\pdfoutput=1
\documentclass[9pt,sigconf]{acmart}

\usepackage[english]{babel}
\usepackage{booktabs}
\usepackage{caption}
\usepackage[capitalise,nameinlink,noabbrev]{cleveref}
\usepackage{enumitem}
\usepackage{etoolbox}
\usepackage{graphicx}
\usepackage{subcaption}
\usepackage{tablefootnote}
\usepackage{titlesec}
\usepackage{xspace}

\acmSubmissionID{424}
\setcopyright{acmcopyright}
\copyrightyear{2023}
\acmYear{2023}
\acmDOI{10.1145/3603269.3604848}
\acmConference[SIGCOMM '23]{SIGCOMM}{2023}{New York City, NY}
\acmPrice{15.00}
\acmISBN{979-8-4007-0236-5/23/09}

\setlist[itemize]{nosep, noitemsep, topsep=0pt, leftmargin=1em}
\titlespacing*{\section}{0pt}{1ex}{.5ex}
\titlespacing*{\subsection}{0pt}{1ex}{.5ex}
\titlespacing*{\subsubsection}{0pt}{.75ex}{.5em}
\titlespacing*{\paragraph}{0pt}{.5ex}{.5em}
\titleformat*{\subsection}{\large\bfseries}
\providetoggle{useminted}
\settoggle{useminted}{false}
\iftoggle{useminted}{
\usepackage{minted}
\definecolor{mintedhlcolor}{rgb}{0.95,0.95,1}
\setminted{
    fontsize=\footnotesize,
    baselinestretch=.95,
    frame=lines,
    highlightcolor=mintedhlcolor,
    python3=true,
    xleftmargin=1em,
    linenos,
    numbersep=6pt
}
\setmintedinline{fontsize=\small}
\newcommand{\code}[1]{\mintinline{text}{#1}}
\newenvironment{codeblock}[2]{\begin{listing}[ht]\inputminted[#1]{python}{#2}}{\end{listing}}
}{  % else if useminted is false:
\usepackage{listings}
\usepackage[frozencache]{minted}
\definecolor{codegreen}{rgb}{0,0.6,0}
\lstset{
  basicstyle=\linespread{1}\footnotesize\ttfamily,
  columns=flexible,
  commentstyle=\color{codegreen},
  frame=lines,
  keywordstyle=\color{blue},
}
\newcommand{\code}[1]{\lstinline[basicstyle=\small\ttfamily]|#1|}
\newenvironment{codeblock}[2]{\begin{listing}[ht]\lstinputlisting[language=python]{#2}}{\end{listing}}
}

\hypersetup{breaklinks=true}

\crefformat{section}{#2\S#1#3}
\Crefformat{section}{#2Section~#1#3}
\crefmultiformat{section}{#2\S\S#1#3}{ and~#2#1#3}{, #2#1#3}{, and~#2#1#3}
\Crefmultiformat{section}{#2Sections~#1#3}{ and~#2#1#3}{, #2#1#3}{, and~#2#1#3}
\crefname{figure}{Fig}{Figures}
\Crefname{figure}{Figure}{Figures}
\crefname{codeblock}{Listing}{Listings}
\Crefname{codeblock}{Listing}{Listings}

\newcommand{\nbc}[3]{\textcolor{#3}{\sf\small\textit{#1: #2}}}
\definecolor{color1}{HTML}{2196F3}
\definecolor{color2}{HTML}{9C27B0}
\definecolor{color3}{HTML}{009688}
\definecolor{color4}{HTML}{FF5722}
\definecolor{color5}{HTML}{673AB7}
\definecolor{color6}{HTML}{E91E63}
\newcommand\todo[1]{\textcolor{red}{\sf\small\textit{\textbf{TODO:} #1}}}
\newcommand\lsf[1]{\nbc{Frank}{#1}{color1}}
\newcommand\samyu[1]{\nbc{Samyu}{#1}{color3}}
\newcommand\swang[1]{\nbc{Stephanie}{#1}{color2}}
\newcommand\ekl[1]{\nbc{Eric}{#1}{color2}}
\newcommand\ion[1]{\nbc{Ion}{#1}{color4}}
\renewcommand\todo[1]{}
\renewcommand\lsf[1]{}
\renewcommand\samyu[1]{}
\renewcommand\swang[1]{}
\renewcommand\ekl[1]{}
\renewcommand\ion[1]{}
\newcommand\revision[1]{\textcolor{color4}{#1}}
\renewcommand\revision[1]{#1}

\newcommand{\sys}{{Exoshuffle}\xspace}
\newcommand{\sysshort}{{ES}}

\begin{document}

\begin{abstract}

Shuffle is one of the most expensive communication primitives in distributed data processing and is difficult to scale.
Prior work addresses the scalability challenges of shuffle by building monolithic shuffle systems.
These systems are costly to develop, and they are tightly integrated with batch processing frameworks that offer only high-level APIs such as SQL.
New applications, such as ML training, require more flexibility and finer-grained interoperability with shuffle. They are often unable to leverage existing shuffle optimizations.

We propose an extensible shuffle architecture.
We present \sys, a library for distributed shuffle that offers competitive performance and scalability as well as greater flexibility than monolithic shuffle systems.
We design an architecture that decouples the shuffle control plane from the data plane without sacrificing performance.
We build \sys on Ray, a distributed futures system for data and ML applications, and demonstrate that we can: (1) rewrite previous shuffle optimizations as application-level libraries with an order of magnitude less code, (2) achieve shuffle performance and scalability competitive with monolithic shuffle systems, and break the CloudSort record as the world's most cost-efficient sorting system, and (3) enable new applications such as ML training to easily leverage scalable shuffle.

\end{abstract}




\newcommand{\papertitle}{\sys: An Extensible Shuffle Architecture}
\title[\papertitle]{\papertitle}

\author{Frank Sifei Luan}
\orcid{0001-8709-6823}
\affiliation{\institution{UC Berkeley}\city{Berkeley}\state{CA}\country{USA}}

\author{Stephanie Wang}
\affiliation{\institution{UC Berkeley and Anyscale}\city{Berkeley}\state{CA}\country{USA}}
  
\author{Samyukta Yagati}
\affiliation{\institution{UC Berkeley}\city{Berkeley}\state{CA}\country{USA}}
  
\author{Sean Kim}
\affiliation{\institution{UC Berkeley}\city{Berkeley}\state{CA}\country{USA}}
  
\author{Kenneth Lien}
\affiliation{\institution{UC Berkeley}\city{Berkeley}\state{CA}\country{USA}}
  
\author{Isaac Ong}
\affiliation{\institution{UC Berkeley}\city{Berkeley}\state{CA}\country{USA}}
  
\author{Tony Hong}
\affiliation{\institution{UC Berkeley}\city{Berkeley}\state{CA}\country{USA}}
  
\author{SangBin Cho}
\affiliation{\institution{Anyscale}\city{San Francisco}\state{CA}\country{USA}}
  
\author{Eric Liang}
\affiliation{\institution{Anyscale}\city{San Francisco}\state{CA}\country{USA}}
  
\author{Ion Stoica}
\affiliation{\institution{UC Berkeley}\city{Berkeley}\state{CA}\country{USA}}

\renewcommand{\shortauthors}{Luan et al.}

\maketitle

\section{Introduction}
\label{sec:intro}

Shuffle is a fundamental operation in distributed data processing systems.
\revision{It refers to the all-to-all data transfer from mappers to reducers in a MapReduce-like system~\cite{mapreduce}.}
Shuffle is one of the most expensive communication primitives in these systems and is difficult to scale.
Scaling shuffle requires efficiently and reliably moving a large number of small blocks from each mapper to each reducer across memory, disk, and network. It requires both high I/O efficiency, and robustness to failures and data skew. Furthermore, as the data size increases, the number of shuffle blocks grows quadratically, making shuffle the most costly operation in some workloads. 

The difficulty of scaling shuffle has inspired many solutions from both the industry and research community.
These shuffle implementations improve the performance and reliability of large-scale shuffle by optimizing I/O in different storage environments, such as HDD, SSD and disaggregated storage~\cite{sailfish,riffle,magnet,awsglueshuffle}.
Since performance at scale is a priority, these prior solutions are built as monolithic shuffle systems from scratch using low-level system APIs.
However, these systems are costly to develop and integrate.
For example, cloud providers each have to build proprietary services to support shuffle on their own storage services~\cite{awsglueshuffle,alibaba-emr,google-dataflow-shuffle}.
Magnet, a push-based shuffle system for Spark~\cite{magnet}, took 19 months between publication and open-source release in the Spark project~\cite{spark-3-2-0} because it required significant changes to system internals~\cite{spark32915}.

Furthermore, existing shuffle systems only work with batch processing frameworks, which offer high-level abstractions such as SQL or dataframe APIs. Most are synchronous in nature: the results are available only after the entire shuffle operation completes.
This poses challenges for applications that require \emph{fine-grained} integration with the shuffle operation to improve their performance by processing data as it is being shuffled, i.e., pipeline data processing with the shuffle operation.
For example, ML training often requires repeatedly shuffling the training dataset between epochs to improve learning quality~\cite{randomreshuffling,sgd-shuffle}.
Doing this efficiently requires fine-grained pipelining between shuffle and training: ML trainers should consume partial shuffle outputs as soon as they become ready.
Today's ML developers are faced with two undesirable choices: (1) they either rebuild shuffle from scratch, once again dealing with the performance challenges of large-scale shuffle, or (2) interface with existing shuffle systems through the synchronous APIs: the shuffle results can only be consumed after all partitions are materialized, leaving pipelining opportunities on the table.

\begin{figure}[t]
  \centering
  \begin{subfigure}[b]{0.5\linewidth}
    \centering
    \includegraphics[bb=0 0 720 405,width=\textwidth,trim=0.5cm 4.5cm 15.5cm 4.5cm, clip]{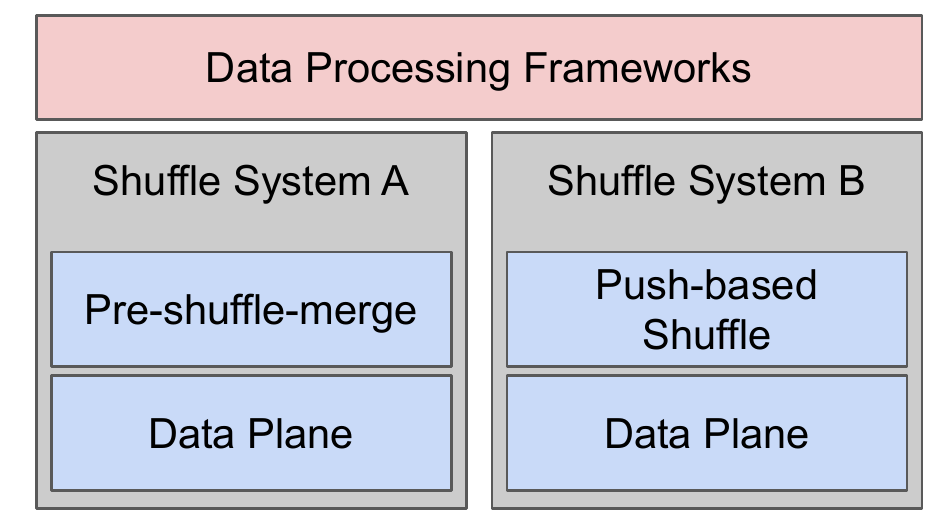}
    \subcaption{Monolithic shuffle systems.}
    \label{fig:shuffle-arch:monolithic}
  \end{subfigure}%
  \begin{subfigure}[b]{0.5\linewidth}
    \centering
    \includegraphics[bb=0 0 720 405,width=\textwidth,trim=11.2cm 4.5cm 4.8cm 4.5cm, clip]{figures/exoshuffle.pdf}
    \subcaption{\sys.}
    \label{fig:shuffle-arch:exoshuffle}
  \end{subfigure}
  \vspace*{-1ex}
  \label{fig:shuffle-arch}
  \caption{\sys builds on an extensible architecture. Shuffle as a library is easier to develop and more flexible to integrate with applications. The data plane ensures performance and reliability.
  }
\end{figure}

To simplify the development of new shuffle optimizations targeting different environments, and to provide fine-grained pipelining for new applications, we propose an \emph{extensible} architecture for distributed shuffle that enables flexible, efficient, and scalable implementations.
Unlike previous solutions built as monolithic systems~(\cref{fig:shuffle-arch:monolithic}),
we propose building distributed shuffle as a library~(\cref{fig:shuffle-arch:exoshuffle}).
Such an architecture allows: (1) shuffle builders to easily develop and integrate new shuffle designs for new environments, and (2) a broader set of applications to leverage scalable shuffle in a more flexible manner.

How can we implement shuffle at the application level (as a library) while providing high performance? To answer this question, we first identify the optimizations in past shuffle systems that are key to performance and reliability.
(1) Coordination: managing the timing and placement of mapper and reducer tasks, and implementing optimizations such as merging intermediate shuffle blocks.
(2) Efficient data transfer: pipelining I/O with computation to maximize throughput, and spilling data to disk to accommodate larger-than-memory datasets.
(3) Fault tolerance: guaranteeing data is reliably transferred to reducers via retries or replication.

Our key observation is that we can split these optimizations between a control and a data plane.
Optimizations for coordinating shuffle are implemented by
the \emph{control plane} at the application layer, while the \emph{data plane} provides efficient data transfer and fault tolerance at the system layer. This enables developers to easily implement a variety of shuffle solutions at the application layer, while having the underlying system handle efficient data transfer and fault tolerance.

The next question is what interface should the data plane provide to the application.
Our answer is~\emph{distributed futures},
an extension of RPC that allows referencing data objects in distributed memory.
It allows the caller of a remote task to pass objects \emph{by reference}, regardless of their physical locations, thus decoupling remote task invocations from physical data transfers, the latter implemented by the data plane.
Distributed futures can also be passed before the data object is created, allowing the system to parallelize remote calls and pipeline data transfer with task execution.
We show that this abstraction can express a variety of shuffle algorithms, including dynamic strategies to handle data skew and stragglers (\cref{sec:shuffle}).

Although many distributed futures implementations exist, none of the these systems have been able to match the scale and performance of a monolithic shuffle system.
CIEL~\cite{ciel} is the first to show MapReduce programs can be implemented using distributed futures, but it lacks an in-memory object store which is crucial for efficient pipelining and data transfers.
Dask~\cite{dask}, another distributed futures-based dataframe system, supports in-memory objects but cannot scale beyond hundreds of GBs~(\cref{sec:eval:dask}).
Previous versions of Ray~\cite{ray} support shuffle within the capacity of its distributed shared memory object store, \revision{but lack disk spilling mechanisms and therefore do not support out-of-core processing.}

In this work, we extend Ray with the necessary features to support large-scale shuffle (\cref{sec:arch}). These include:
(1) locality scheduling primitives to enable colocating tasks to better exploit shuffle data locality;
(2) a full distributed memory hierarchy with disk spilling and recovery;
(3) asynchronous object fetching to pipeline task execution with disk and network I/O.
We present \sys, a flexible and scalable library for distributed shuffle built on top of Ray.
We demonstrate the advantages of this extensible shuffle architecture by showing that (\cref{sec:eval}):
\begin{itemize}
    \item A variety of previous shuffle optimizations can be written as distributed futures programs in \sys, with an order of magnitude less code.
    \item The \sys implementations of these shuffle optimizations match or exceed the performance of their monolithic counterparts.
    \item \sys can scale to 100\,TB, outperforming Spark and Magnet by $1.8\times$, and breaking the CloudSort record as the world's most cost-efficient sorting system.
    \item \sys can easily integrate with a diverse set of applications such as distributed ML training, improving end-to-end training throughput by $2.4\times$.
\end{itemize}

\begin{figure*}[t]
  \centering
  \begin{subfigure}[b]{0.16\linewidth}
    \centering
    \includegraphics[bb=0 0 720 405,width=\linewidth,trim=0cm 3.5cm 18cm 2.5cm, clip]{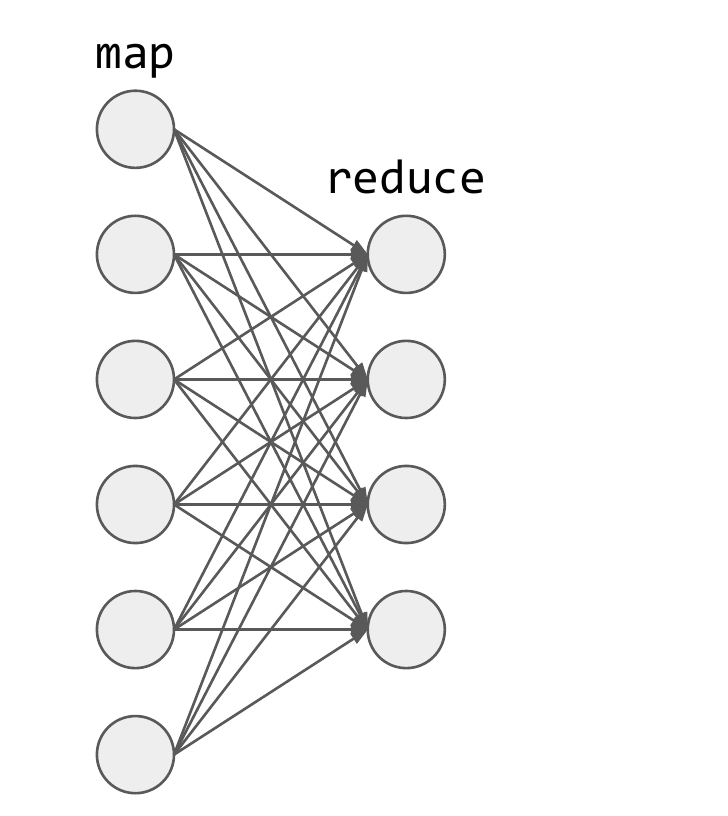}
    \subcaption{``Simple'' shuffle~\cite{mapreduce}.}
    \label{fig:shuffle-dags:naive}
  \end{subfigure}%
  \begin{subfigure}[b]{0.18\linewidth}
    \centering
    \includegraphics[bb=0 0 720 405,width=\linewidth,trim=8cm 3.5cm 10cm 2.5cm, clip]{figures/shuffle-dags.pdf}
    \subcaption{Pre-shuffle merge~\cite{riffle}.}
    \label{fig:shuffle-dags:riffle}
  \end{subfigure}%
  \begin{subfigure}[b]{0.22\linewidth}
    \centering
    \includegraphics[bb=0 0 720 405,width=.82\linewidth,trim=17cm 3.5cm 1cm 2.5cm, clip]{figures/shuffle-dags.pdf}
    \subcaption{Push-based shuffle~\cite{magnet,cosco,NADSort}.}
    \label{fig:shuffle-dags:magnet}
  \end{subfigure}%
  \begin{subfigure}[b]{0.44\linewidth}
    \centering
    \includegraphics[bb=0 0 720 405,width=\linewidth,trim=0.3cm 2.5cm 2cm 2cm, clip]{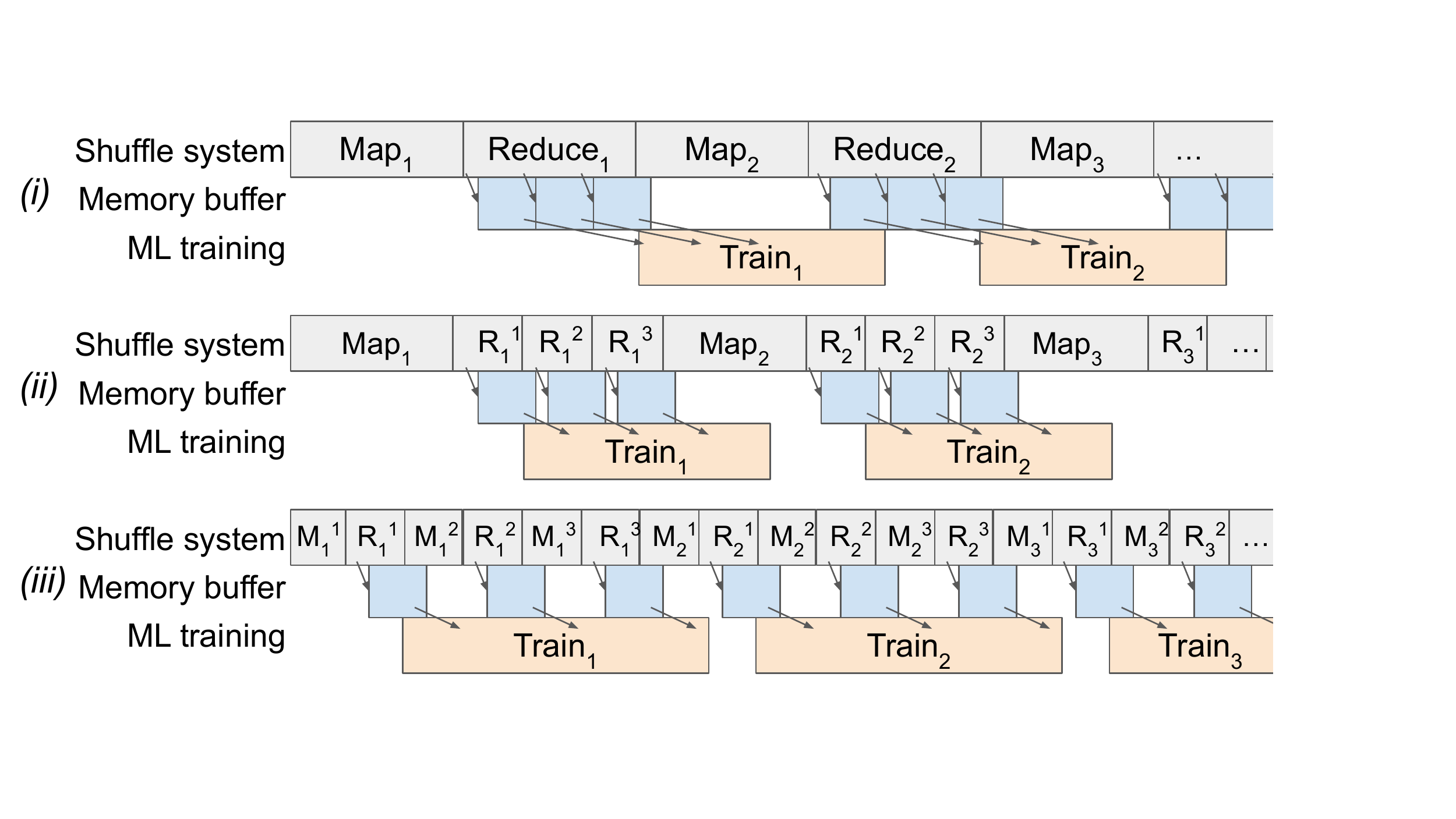}
    \subcaption{ML training pipeline.}
    \label{fig:shuffle-dags:ml}
  \end{subfigure}%
  \caption{Shuffle algorithms for various applications. \sys uses distributed futures to execute these DAGs.}
  \label{fig:shuffle-dags}
\end{figure*}
\section{Motivations}
\label{sec:background}

In this section, we overview two lines of previous work in building shuffle systems to illustrate the challenges in simultaneously achieving shuffle scalability and flexibility.

\subsection{Shuffle Systems}
\label{sec:background:sort}

\begin{table}[ht]
  \begin{tabular}{ll}
    \toprule
    Storage target
    &
    Shuffle systems
    \\
    \midrule
    Hard disk
    &
    Sailfish~\cite{sailfish}, Riffle~\cite{riffle}, Magnet~\cite{magnet} \\
    SSD
    &
    Zeus~\cite{zeus}
    \\
    Cloud storage
    & Alibaba E-MapReduce Shuffle~\cite{alibaba-emr}, \\
    & AWS Glue Shuffle~\cite{awsglueshuffle}, \\
    & Google Cloud Dataflow Shuffle~\cite{google-dataflow-shuffle} \\
  \bottomrule
\end{tabular}
  \caption{Different shuffle systems are built to optimize shuffle for deployment in different storage environments.}
  \label{tab:shuffle-systems}
\end{table}

In a MapReduce operation with $M$ map tasks and $R$ reduce tasks, shuffle creates $M\times R$ intermediate blocks.
Each of these blocks must be moved across memory, disk, and network.
As the number of tasks grow, the number of blocks increases and the block size decreases both quadratically.
At terabyte scale, this can result in hundreds of millions of very small blocks.
This creates great challenges for I/O efficiency, especially for hard drives with low IOPS limits.
Many shuffle systems have been built to optimize I/O efficiency in different storage environments. \Cref{tab:shuffle-systems} shows an incomplete list of these systems, grouped by their target storage environments.

Previous I/O optimizations fall under two general categories:
(1) reducing the number of small and random I/O accesses by \emph{merging} intermediate blocks into larger ones at various stages~\cite{riffle,cosco,magnet}~(\Cref{fig:shuffle-dags:riffle,fig:shuffle-dags:magnet}), and (2) using \emph{pipelining} to overlap I/O with execution~\cite{magnet,sparkshufflemagic}.
For example, \emph{push-based shuffle}~\cite{NADSort,ishuffle} involves pushing intermediate outputs directly from the mappers to the reducers, allowing network and disk I/O to be overlapped with map execution, and optionally merging results on the reducer~(\Cref{fig:shuffle-dags:magnet}) to improve disk write efficiency~\cite{magnet,cosco}.

While these solutions can improve throughput, they also come with high development cost.
Each new operation, such as reduce-side merge, requires building additional protocols for managing block transfers. %, as well as different tradeoffs.
However, although the ideas may be system-agnostic, the physical artifacts are often tightly integrated with proprietary storage systems, making them difficult to port to open-source frameworks.
For example, many cloud providers build proprietary shuffle services to work with their own disaggregate storage offerings~\cite{awsglueshuffle,alibaba-emr,google-dataflow-shuffle}; meanwhile, Magnet~\cite{magnet} is open-sourced as part of Spark but has yet to support disaggregated storage.

Furthermore, large-scale shuffle systems often come with more complicated deployment models.
They are often deployed as auxiliary services to existing data processing systems.
Shuffle services decouple block lifetimes from task executors to minimize interruptions upon executor failures~\cite{sparkpetabyteblog}, which are more frequent in large clusters.
Shuffle services are also used to coordinate more sophisticated shuffle protocols, such as push-based shuffle and reduce-side merge~\cite{magnet}.
However, because these shuffle services are only necessary at very large scale, they are not enabled by default in systems like Spark and require a separate deployment process.

Thus, while there has been significant innovation in new shuffle designs, few of these are widely deployed.
Furthermore, it is difficult for an application to choose on the fly whether to use a particular shuffle algorithm; it requires both a priori knowledge of the application scenario and potentially an entirely different system deployment.

\subsection{Random Shuffle in ML Training Pipelines}
\label{sec:background:ml}

While much of the existing shuffle literature has focused on large-scale batch processing, there is also a need for performant shuffle in other application scenarios, such as online aggregation~\cite{mapreduceonline} and pipelining with more complex applications.
An example of the latter is the \emph{random shuffle} operation commonly used in machine learning training jobs.
Note that by random shuffle, we mean the application-level transform that randomly permutes the rows of a dataset, rather than the generic system-level shuffle that is used to execute MapReduce applications.

To improve model convergence in deep learning, it is common practice to randomly shuffle the training dataset before feeding into GPU trainers to avoid bias on the order of the data~\cite{randomreshuffling,sgd-shuffle}.
To minimize GPU pauses, the shuffle should be pipelined with the training execution~(\Cref{fig:shuffle-dags:ml}).
Furthermore, it is desirable for developers to be able to trade off between performance and accuracy: they might wish to run shuffle in a smaller window to reduce training latency, at the cost of overall end model accuracy.

These differences make it difficult for ML pipelines to directly leverage existing monolithic shuffle systems.
Systems like Hadoop and Spark are highly optimized for global shuffle operations, but are not designed to pipeline the shuffle with downstream executions: shuffle results cannot be read until the full shuffle is complete~\cite{mapreduceonline}.
The results must be written out to an external store before they can be read by the training workers (\Cref{fig:shuffle-dags:ml}$i$).
However, this leads to either high memory footprint, as it requires holding an additional copy of the dataset, or higher I/O overhead, if the shuffled data is written to disk before transfer to the GPU.

\emph{Fine-grained pipelining} can improve efficiency.
\Cref{fig:shuffle-dags:ml}$ii$ shows an example in which the reduce tasks for a particular epoch are pipelined with the training computation.
This allows results to be used as they become available while limiting memory footprint to a single partition.
Alternatively, the application can also choose to shuffle the dataset in windows (\Cref{fig:shuffle-dags:ml}$iii$), improving pipelining at the cost of accuracy.
Unfortunately, existing shuffle systems are not built for such fine-grained pipelining, and most big data systems that offer high-performance shuffle use an execution model that is incompatible with deep learning systems~\cite{bigdl}.

\revision{
Instead, ML training frameworks often end up re-implementing shuffle within specialized data loaders and thus run into known problems that have been solved by traditional shuffle systems.
Typically, data loaders are implemented with a pool of CPU-based workers colocated with the GPU trainers~\cite{tfdata,torchdata,petastorm}.
Each worker loads a partition of the dataset from storage (e.g., Amazon S3), preprocesseses it, and feeds the resulting data into the colocated trainers.
To support random shuffle, the workers may read a random partition of the dataset on each epoch.
However, to improve I/O efficiency, data must still be read in batches.
Thus, to de-correlate data within the same batch, workers further shuffle the data by mixing records within a fixed-size local memory buffer.
This effectively ties the shuffle window size to the size of the memory buffer.
Setting the buffer size too large results in out-of-memory errors and poor pipelining, but if the buffer is too small, data de-correlation may be insufficient.
In \Cref{sec:eval:ml-training}, we demonstrate how \sys can bring distributed shuffle optimizations to ML training applications, achieving both high performance and flexibility.
}

\section{Shuffle with Distributed Futures} \label{sec:shuffle}
For distributed futures to serve as an intermediate abstraction layer for shuffle, they should: (1) abstract out the common implementation details of different shuffle implementations, (2) be general enough to allow heterogeneous end applications to interface with the shuffle library, and (3) provide the same performance and reliability as monolithic shuffle systems.
This narrow waist for distributed shuffle would enable both faster development for new shuffle implementations and extensibility to new application use cases.

Monolithic shuffle systems use messaging primitives, like RPC, as an intermediate abstraction layer. 
RPC is both general-purpose and high-performance, but it is too low-level to be a useful intermediate layer for shuffle.
Integrating push-based shuffle into Spark, for example, required 1k+ LoC for the RPC layer changes alone~\cite{spark32915}. 
Much of this development effort lies in implementing new inter-task protocols for data transfer and integrating them alongside existing ones.

In contrast, distributed futures decouple the shuffle control plane from the data plane. 
This abstraction enables different shuffle libraries to share a common data plane.
Optimizations like push-based shuffle can be implemented in an order of magnitude less code as a result (\cref{sec:eval:complexity}).

In this section, we show how  to express previous shuffle optimizations and application-specific shuffle variants as application-level programs with the distributed futures API.
These simplified examples capture the logical execution DAG of the shuffle.
\Cref{sec:arch} describes the physical execution of these programs and the details in achieving performance parity with monolithic shuffle systems.

\subsection{The Distributed Futures API} \label{sec:shuffle:api}

A distributed futures program invokes remote functions, known as \emph{tasks}, that execute and return data on a remote node.
When calling a remote function,
the caller immediately gets a distributed future that represents the \emph{eventual} return value.
The future is ``distributed'' because the return value may be stored anywhere in the cluster, e.g., at the node where the task executes.
This avoids copying return values back to the caller, which can become expensive for large data.

The caller can make use of a distributed future in two ways.
First, it can create a DAG by passing a distributed future as an argument to another task.
The system ensures that the dependent task runs only after all of its arguments are computed.
Note that the caller can specify such dependencies before the value is computed and that the caller need not see the physical values.
This gives the system control over parallelism and data movement, e.g., pipelining task execution with dependency fetching for other tasks, and allows the caller to manipulate data larger than local memory.
Second, the caller can get the value of a distributed future using a \code{get} call, which fetches the value to the caller's local memory.
This is useful when consuming the output of a shuffle, as it allows the caller to pipeline its own execution with the shuffle.
The caller can additionally use a \code{wait} call, which blocks until a set of tasks complete (without fetching the return values), for synchronization and for avoiding scheduling too many concurrent tasks.

\subsection{Expressing Shuffle with Distributed Futures}

We demonstrate how these APIs can be used to express various shuffle optimizations~(\cref{fig:shuffle-dags}) as application-level programs.
We use Ray's distributed futures API for Python~\cite{ray} for illustration.
The \code{@ray.remote} annotation designates remote functions, and the \code{.remote()} operator invokes tasks.

\begin{codeblock}{}{listings/shuffle-algorithms.py}
\caption{Shuffle algorithms as distributed futures programs.}
\label{lst:shuffle-algorithms}
\end{codeblock}

\subsubsection{Simple Shuffle} \label{sec:shuffle:impl:naive}

In \Cref{lst:shuffle-algorithms}, \code{simple_shuffle} shows a straightforward implementation of the MapReduce paradigm illustrated in \Cref{fig:shuffle-dags:naive}.
The shuffle routine takes a \code{map} function that returns a list of map outputs, and a \code{reduce} function that takes a list of map outputs and returns a reduced value. \code{M} and \code{R} are the numbers of map and reduce tasks respectively.
The two statements produce the task graph shown in \Cref{fig:shuffle-dags:naive}. Note that the \code{.remote()} calls are non-blocking, so the entire task graph can be submitted to the system without waiting for any one task to complete.

This is effectively \emph{pull-based shuffle}, in which shuffle blocks are \emph{pulled} from the map workers as reduce tasks progress.
Assuming a fixed partition size, the total number of shuffle blocks grows quadratically with the total data size.
\Cref{sec:eval:performance} shows empirical evidence of this problem: as the number of shuffle blocks increases, the performance of the naive shuffle implementation drops due to decreased I/O efficiency.
Prior work~\cite{riffle,magnet,cosco} have proposed solutions to this problem, which we study and compare next.

\subsubsection{Pre-Shuffle Merge} \label{sec:shuffle:impl:riffle}

Riffle~\cite{riffle} is a specialized shuffle system built for Spark. Its key optimization is merging small map output blocks into larger blocks, thereby converting small, random disk I/O into large, sequential I/O before shuffling over the network to the reducers.
The merging factor $F$ is either pre-configured, or dynamically decided based on a block size threshold.
As soon as $F$ map tasks finish on an executor node, their output blocks ($F\times R$) are merged into $R$ blocks, each consisting of $F$ blocks of data from the map tasks.
This strategy, illustrated in \Cref{fig:shuffle-dags:riffle}, is implemented in \Cref{lst:shuffle-algorithms} (\code{shuffle_riffle}).
The code additionally takes \code{F} as the merging factor, and a merge function which combines multiple map outputs into one.

Riffle's key design choice is to merge map blocks \emph{locally} before they are pulled by the reducers, 
as shown in the highlighted lines.
For simplicity, the code assumes that the first $F$ map tasks are scheduled on the first worker, the next $F$ map tasks on the second worker, etc.
In reality, the locality can be determined using scheduling placement hints or runtime introspection (\cref{sec:arch:scheduling})
\Cref{sec:eval:performance} shows that this implementation of Riffle-style shuffle improves the job completion time over simple shuffle.

\subsubsection{Push-based Shuffle} \label{sec:shuffle:impl:magnet}

Push-based shuffle~(\cref{fig:shuffle-dags:magnet}) is an optimization that pushes shuffle blocks to reducer nodes as soon as they are computed, rather than pulling blocks to the reducer when they are required.
Magnet~\cite{magnet} is a specialized shuffle service for Spark that performs this optimization by merging intermediate blocks on the reducer node before the final reduce stage.
This improves I/O efficiency and data locality for the final reduce tasks.
\code{shuffle_magnet} in \Cref{lst:shuffle-algorithms} implements this design.

\subsubsection{Straggler Mitigation} \label{sec:shuffle:straggler}

Distributed futures enable dynamic task graphs by nature, making it ideal for detecting and reacting to stragglers during runtime.

\paragraph{Speculative Execution}
One way to handle stragglers is through speculative execution.
Tasks that are suspected to be stragglers can be duplicated, and the system chooses whichever result is available first.
This can be accomplished with distributed futures using the \code{ray.wait} primitive, as shown in \cref{lst:speculative}.

\begin{codeblock}{}{listings/speculative-execution.py}
\caption{Mitigating stragglers with speculative execution.}
\label{lst:speculative}
\end{codeblock}

\paragraph{Best-effort Merge}
Shuffle systems including Riffle and Magnet also implement ``best-effort merge'', where a timeout can be set on the shuffle and merge phase~\cite{riffle,magnet}. If some merge tasks are cancelled due to timeout, the original map output blocks will be fetched instead.
This ensures straggler merge tasks will not block the progress of the entire system.
Best-effort merge can be implemented in \sys as shown in \Cref{lst:timeout} using an additional \code{ray.cancel()} API which cancels the execution of a task.
The cancelled task's input, which are the original map output blocks, will then be directly passed to the reducers.
This way, the task graph is dynamically constructed as the program runs, adapting to runtime conditions while still enjoying the benefits of transparent fault tolerance provided by the system.

\begin{codeblock}{}{listings/task-timeout.py}
\caption{Mitigating stragglers via task cancellation.}
\label{lst:timeout}
\end{codeblock}

\subsubsection{Data Skew}
\label{sec:shuffle:skew}

Data skew can be prevented at the data management level using techniques such as key salting, or periodic repartitioning.
However, it is still possible for skews to occur during ad-hoc query processing, especially for those queries involving joins and group-bys.
Data skew during runtime can cause the working set of a reduce task to be too large to fit into executor memory.

Dynamic repartitioning solves this problem by further partitioning a large reducer partition into smaller ones.
This is straightforward to implement since the distributed futures programming model enables dynamic tasks by nature. \Cref{lst:skew} shows that we can recursively split down a reducer's working set until it fits into a predefined memory threshold.

\begin{codeblock}{}{listings/skew.py}
\caption{Dynamic repartitioning for skewed partitions.}
\label{lst:skew}
\end{codeblock}

\subsection{Applications}
\label{sec:shuffle:app}

Because \sys implements shuffle at the application level, it can easily interoperate with other applications.
Here, we demonstrate two example applications that use fine-grained pipelining with shuffle to improve end-to-end performance.
These applications are evaluated in \Cref{sec:eval:app}.

\subsubsection{Online Aggregation with Streaming Shuffle}
\label{sec:shuffle:app:mpo}

Online aggregation~\cite{online-aggregation} is an interactive query processing mode where partial results are returned to the user as soon as some data is processed, and are refined as progress continues.
This is especially useful when the query takes a long time to complete.
Online aggregation is difficult to implement in MapReduce systems because they require all outputs to be materialized before being consumed.
Past work made in-depth modifications to Hadoop and Spark to support online aggregation~\cite{mapreduceonline, g-ola}.

\begin{codeblock}{}{listings/streaming-shuffle-lib.py}
    \caption{Streaming shuffle and pipelined data loading for ML.}
    \label{lst:streaming-shuffle}
\end{codeblock}

Online aggregation is straightforward to implement in \sys without the need to modify the underlying distributed futures system.
\Cref{lst:streaming-shuffle} shows the \code{streaming_shuffle} routine. It requires a modified \code{reduce} function that takes a reducer state and a list of map outputs and returns an updated state, and an \code{aggregate} function which combines the reducer states to produce aggregate statistics.
Shuffle is executed in rounds. At the end of each round, the aggregation function is invoked with the reducer outputs, and will asynchronously print an aggregate statistic (e.g. sum) to the user.
Note that the \sys user can simply swap between \code{simple_shuffle} and \code{streaming_shuffle} to get the semantics they desire.

\subsubsection{Distributed ML Training with Pipelined Shuffle}

\sys also enables fine-grained pipelining for ML training, as illustrated in \Cref{fig:shuffle-dags:ml}. In \Cref{lst:streaming-shuffle}, \code{model_training} shows the code skeleton.
On line 13, the \code{shuffle} function (could be any in \Cref{lst:shuffle-algorithms}) returns a set of distributed futures pointing to reducer outputs.
They are passed immediately to the model trainer while shuffle executes asynchronously.
As soon as a reducer block becomes available, the model trainer acquires it (line 17) and send it to the GPU for training.
This achieves the fine-grained pipelining described in \Cref{fig:shuffle-dags:ml}.

\section{System Architecture}
\label{sec:arch}
\Cref{sec:shuffle} shows how shuffle DAGs can be expressed as distributed futures programs.
However, achieving high performance shuffle also requires a set of critical system facilities.
In this section, we describe the architecture of \sys via a realistic implementation of the push-based shuffle described in \Cref{sec:shuffle:impl:magnet}.
We describe the additional system APIs used by \sys (\cref{sec:arch:scheduling}), and the transparent features provided by the underlying distributed futures implementation (\cref{sec:arch:system-facilities}) that are key to performance.

\subsection{Example: Push-based Shuffle}
\label{sec:arch:example}
\begin{codeblock}{}{listings/two-stage-shuffle-2.py}
    \caption{Implementation of two-stage shuffle.}
    \label{lst:two-stage-shuffle}
\end{codeblock}

\cref{lst:two-stage-shuffle} implements push-based shuffle~(\cref{sec:shuffle:impl:magnet}) for a cluster of \code{NUM_WORKERS} nodes.
The library takes a \code{map} and a \code{reduce} function as input.
The remaining constants are chosen by the library according to the user-specified number of input and output partitions.

Lines 11--25 comprise the map and merge stage, in which map results are shuffled, pushed to the reducer nodes, and merged.
This stage pipelines between CPU (\code{map} and \code{merge} tasks), network (to move data between \code{map} and \code{merge}), and disk (to write out \code{merge} results).
The map and merge tasks are scheduled in rounds for pipelining: Lines 18--19 ensures that there is at most one round of merge tasks executing, and that they can overlap with the following round's map tasks.
Each round submits one merge task per worker node.
Each merge task takes in one intermediate result from each map task from the same round and returns as many merged results as there are reduce partitions on that worker.

Once all map and merge tasks are complete, we schedule all reduce tasks (lines 28--31) and return the distributed future results.
Each reduce task performs a final reduce on all merge results for its given partition.
To minimize unnecessary data transfer, the reduce tasks are co-located with the merge tasks whose results they read.

\subsection{Scheduling Primitives}
\label{sec:arch:scheduling}

For complex applications like distributed shuffle, it is difficult for a general-purpose system to make optimal decisions in every context.
For instance, optimally scheduling a computation DAG on a set of nodes is NP-hard~\cite{dag-scheduling}.
It is therefore more robust to allow the application or library developer to apply domain-specific knowledge to achieve better performance.

By default, Ray provides a two-level distributed scheduler that balances between bin-packing vs. load-balancing~\cite{ray}.
This is sufficient for map and reduce tasks in simple shuffle, as these can be executed anywhere in the cluster.
However, more advanced shuffle strategies~(\cref{sec:shuffle:impl:magnet,sec:shuffle:impl:riffle}) require more careful placement and scheduling of tasks to improve performance.
In this section, we describe the additional APIs designed to give the shuffle library more control over the physical execution of the shuffle DAG.

\begin{figure*}[t]
  \centering
  \begin{subfigure}[b]{0.5\textwidth}
    \centering
    \includegraphics[width=.75\textwidth,trim=1.7cm 0cm 3.8cm 4cm, clip]{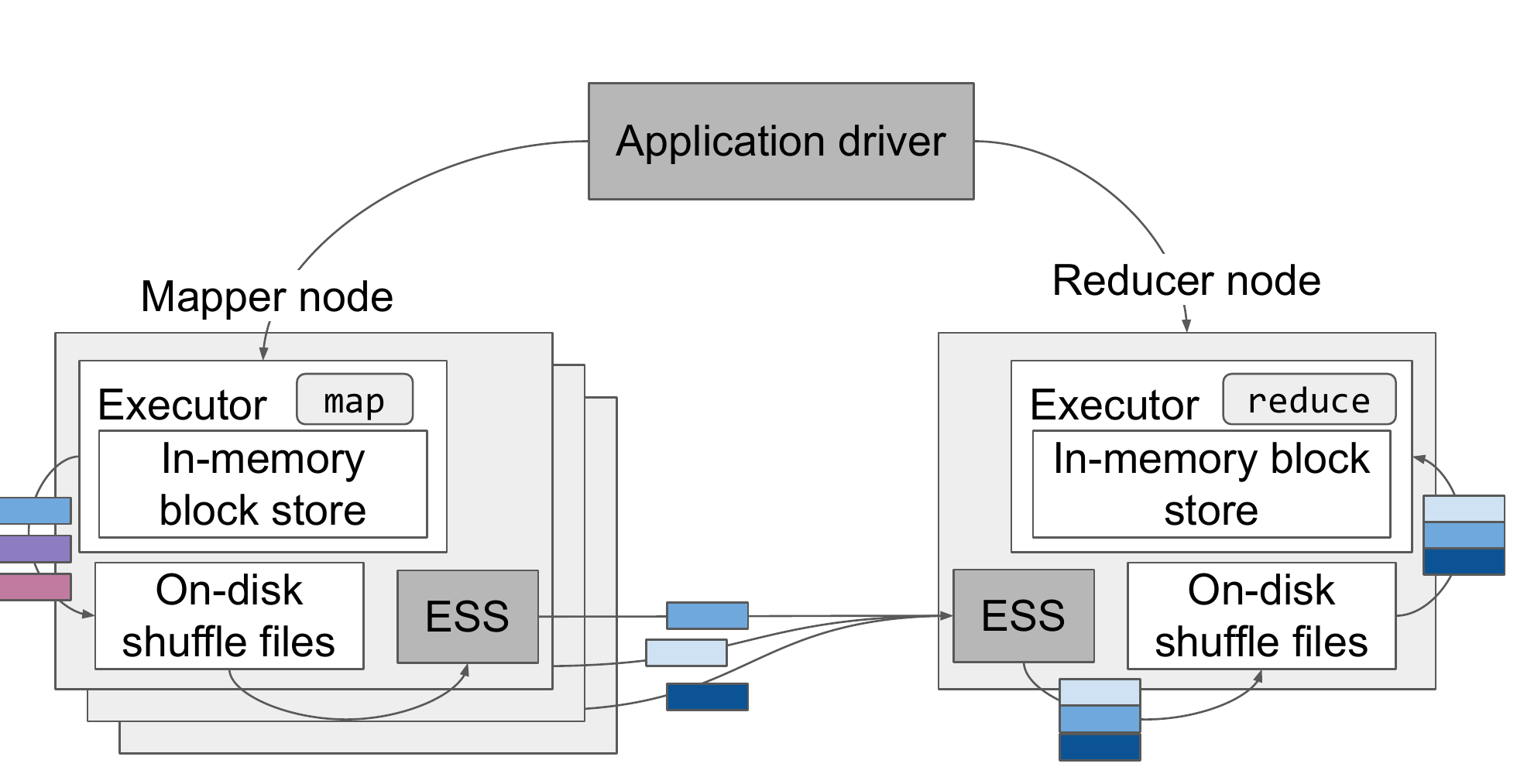}
    \subcaption{Example of a monolithic shuffle architecture.}
    \label{fig:arch:magnet}
  \end{subfigure}%
  \begin{subfigure}[b]{0.5\textwidth}
    \centering
    \includegraphics[width=.75\textwidth,trim=1.3cm 0cm 3.6cm 4cm, clip]{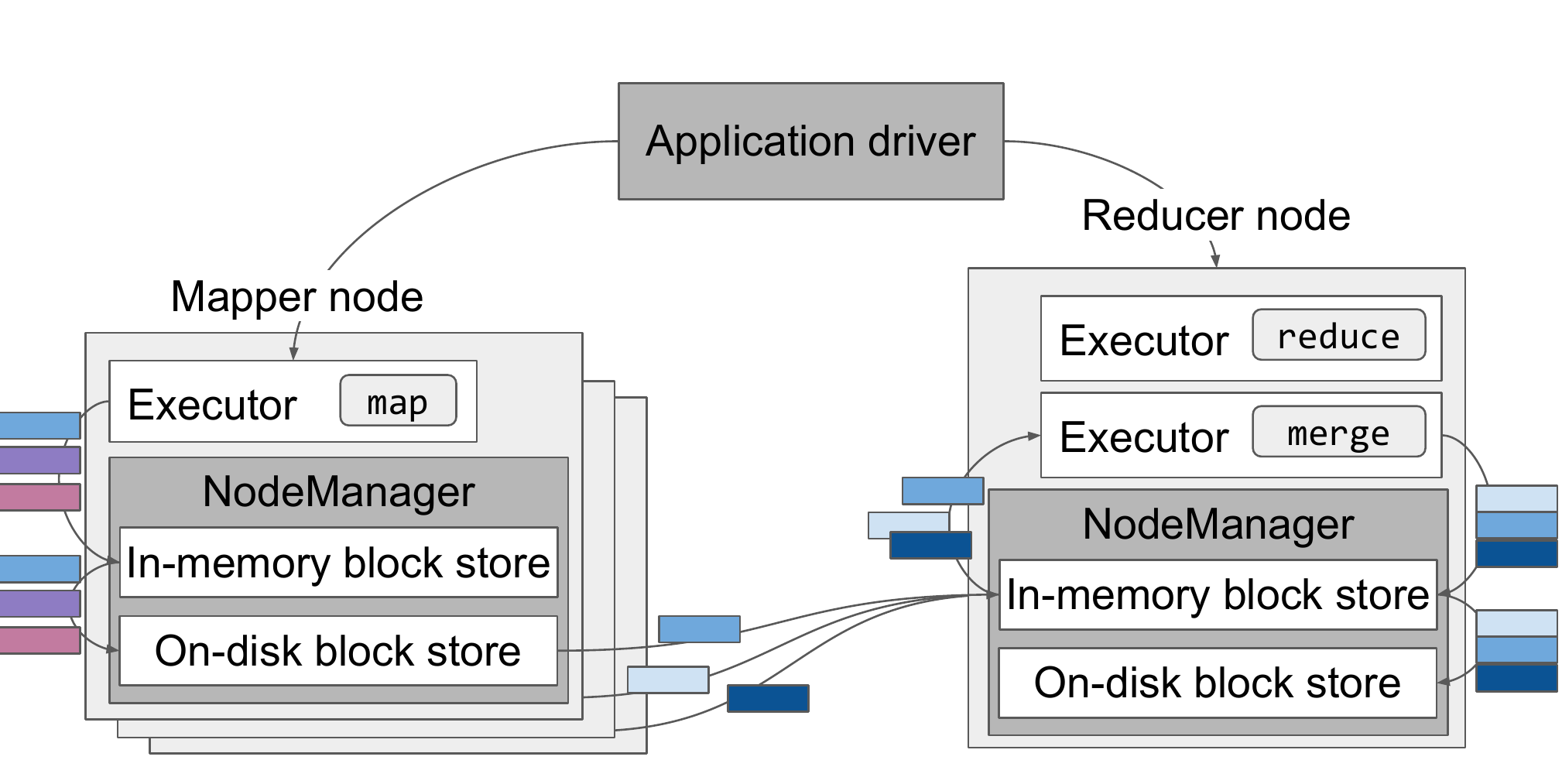}
    \subcaption{\sys.}
    \label{fig:arch:ray}
  \end{subfigure}
  \caption{Comparing a monolithic vs. application-level shuffle architecture. (a) implements all coordination and block management through an external shuffle service on each node, in this case implementing the Magnet shuffle strategy (\cref{sec:shuffle:impl:magnet}). (b) shows the same shuffle strategy but implemented as an application on a generic distributed futures system.}
  \label{fig:arch}
\end{figure*}

\subsubsection{Scheduling for Data Locality}

Ray provides automatic locality-based scheduling when possible.
For example in \Cref{lst:two-stage-shuffle}, lines 28--31, Ray automatically schedules the reduce tasks on the workers on which the upstream merge results reside.
In some other cases, hints must be provided to the system to achieve better data locality.
For example, a group of merge tasks must be colocated with the downstream reduce task, but this is impossible for the system to determine because the reduce task's dependency is not known to the system yet.
To handle this problem, we introduce \emph{node-affinity scheduling} in Ray, which allows the application to pin tasks to a particular node.
For example, \Cref{lst:two-stage-shuffle} uses this in line 23 to colocate merge tasks for the same reducer.
Node affinity is soft, meaning that Ray will choose another suitable node if the specified node fails.

\subsubsection{Scheduling for Task Pipelining}

The map and merge tasks should be pipelined to allow map results to be shuffled concurrently with map execution.
This task-level pipelining is challenging for a distributed futures system to determine automatically:
Too many concurrent map tasks will reduce resources available to downstream merge tasks, and scheduling the wrong set of map and merge tasks concurrently prevents map outputs from being consumed directly by merge tasks, resulting in unnecessary disk writes. 
The shuffle library is better placed to determine that it should apply backpressure by limiting the number of concurrent map and merge tasks.
The library can also determine that a round of merge tasks should be executed concurrently with the following round of map tasks.
\sys achieves this with the \code{wait} API (\cref{lst:two-stage-shuffle}, line 19), which blocks until a task completes.

\subsubsection{Controlling Redundancy with Reference Counting}
\label{sec:arch:scheduling:refs}
Distributed futures are reference-counted in Ray.
While an object reference is in scope, Ray attempts to ensure its value exists in the cluster.
By selecting which references to keep or drop, the shuffle library can make tradeoffs between reducing write amplification and improving data redundancy.
For example, line 25 of \Cref{lst:two-stage-shuffle} deletes the intermediate map results from the current round.
This reduces write amplification, as the map results can be immediately dropped from memory without spilling to disk, but requires additional re-execution upon failure.
Alternatively, the shuffle library can instead keep the intermediate references, resulting in additional disk writes but improved data redundancy.

\subsection{Transparent System Facilities}
\label{sec:arch:system-facilities}

The actual data transfer, or \emph{shuffle}, is managed by the distributed futures system according to the application specifications.
For example in \Cref{lst:two-stage-shuffle}, lines 23--25 specifies that one column of the distributed futures in \code{map_results} should be sent to one merge task.
This prompts the data plane to transfer the corresponding physical data to the \code{merge} task's location.
In this section, we describe the transparent storage and I/O mechanisms provided by the distributed futures system to facilitate this data movement.

\subsubsection{Shared Memory Object Store}
\label{sec:arch:system:obj-store}
Previous monolithic shuffle systems implement distributed coordination via an \emph{external shuffle service}, a specialized process deployed to each node that orchestrates block transfers~(\cref{fig:arch:magnet}).
This process is external to the executors, decoupling block transfers from map and reduce task execution.
In \sys, we replace this service with a generic node manager that is responsible for both in-memory and spilled objects~(\cref{fig:arch:ray}).

We build on Ray's shared memory object store~\cite{ray} for immutable objects. Each node manager hosts a shared memory object store shared by all executors on that node~(\cref{fig:arch:ray}).
This decouples executors from blocks: once a task's outputs are stored in its local object store, the node manager manages the block.
This keeps executors stateless and allows them to execute other tasks or exit safely while the node manager coordinates block movement.
Shared memory enables \textit{zero-copy} reads of object data on the same node, which avoids CPU and memory overhead.
By making objects immutable, we also avoid consistency concerns between object copies.

Next, we describe extensions to the original Ray architecture~\cite{ray} made in this work that improves pipelining disk and network I/O with task execution.
These improvements are made at the system level without knowledge of the application-level shuffle semantics, and thus can benefit a wide range of data-intensive applications.

\subsubsection{Pipelined Object I/O}
\label{sec:arch:memory}

\paragraph{Object Allocation and Fetching.}
\label{sec:arch:memory:allocation}
There are two categories of object memory allocations: new objects created for task returns (e.g., \code{map} task outputs), and copies of objects fetched remotely as task arguments (e.g., \code{merge} task inputs). The memory subsystem queues and prioritizes object allocations to ensure forward progress while keeping memory usage bounded to a limit.
This is critical for reducing thrashing within the object store, caused by requesting objects for too many concurrent requests, while leaving sufficient heap memory for task executors.

All memory allocations on a Ray worker node go into an allocation queue for fulfillment. If there is spare memory, the allocation is fulfilled immediately. 
Otherwise, requests are queued until the spilling process or garbage collection frees up enough memory. 
If memory is still insufficient,
Ray falls back to allocating task output objects on the filesystem to ensure liveness.
Spare memory besides the memory allocated to executing task arguments and returns is used to fetch the arguments of queued tasks.
This enables pipelining between execution and I/O, i.e. restoring objects from disk or fetching objects over the network.
For example, at line 28 in ~\cref{lst:two-stage-shuffle}, all merge results are already spilled to disk and all reduce tasks are submitted at once.
While earlier reduce tasks execute, the system uses any spare memory to restore merge results for the next round of reduce tasks from disk.

\paragraph{Object Spilling.}
\label{sec:arch:memory:spill}
Object spilling is transparent, so the application need not specify if or when it should occur. 
When the memory allocation subsystem has backlogged requests, the spilling subsystem migrates referenced objects to disk to free up memory.
When a spilled object's data is required locally for a task, e.g., because it is the argument of a queued task, the node manager copies it back to memory as described above. %This is handled similarly to allocating for a local copy of an existing object on a remote node.
When requested by a remote node, the spilled object is streamed directly from disk across the network to the remote node manager.
To improve I/O efficiency, Ray coalesces small objects into larger files before writing to the filesystem.

\subsubsection{Fault Tolerance}
\label{sec:arch:ft}

\sys relies on lineage reconstruction for distributed futures to recover objects lost to node failures~\cite{ownership}, a similar mechanism to previous shuffle systems~\cite{mapreduce,spark-rdd}.
In Ray, the application driver stores the object lineage and resubmits tasks as needed upon failure.
This process is transparent to \sys, which runs at the application level. Still, \sys can use object references (\cref{sec:arch:scheduling:refs}) to specify reconstruction or eviction for specific objects.

Executor process failures are much more common than node failures.
If reconstruction is required each time an executor fails, it can impede progress~\cite{sparkpetabyteblog,magnet}.
Many previous shuffle systems use an external shuffle service to ensure map output availability in the case of executor failures or garbage collection pauses.
Similarly, in \sys, executor process failures do not result in the loss of objects, because the object store is run inside the node manager as a separate process.

More sophisticated shuffle systems require additional protocols such as deduplication to ensure fault tolerance~\cite{cosco}.
Distributed futures prevent such inconsistencies because they require objects to be immutable, task dependencies to be fixed, and tasks to be idempotent.

To reduce the chance of data loss, some shuffle system uses on-disk~\cite{magnet} or in-memory~\cite{cosco} replication of intermediate blocks to guard against single node failures.
In Ray, objects are spilled to disk and transferred to remote nodes where they are needed, which also results in multiple copies as long as the object is in scope.
The application can also disable this optimization by deleting its references to the object~(\cref{lst:two-stage-shuffle}, L25).
In the future, we could allow the application to more finely tune the number of replicas kept, e.g., by passing this as a parameter during task invocation.

\section{Evaluation} \label{sec:eval}

We study the following questions in the evaluation:
\begin{itemize}
\item Can \sys libraries achieve performance and scalability competitive with monolithic shuffle systems? (\cref{sec:eval:performance})
\item Is it easier to implement shuffle optimizations in \sys? (\cref{sec:eval:complexity})
\item What benefits does \sys provide for applications, including CloudSort, online aggregation and ML training? (\cref{sec:eval:app})
\item How do the features in the distributed futures backend contribute to \sys performance? (\cref{sec:eval:mb})
    % \item What tradeoffs do applications want to make with shuffle implementations? Can \sys help them to make these tradeoffs and achieve good end-to-end performance? (\cref{sec:eval:app-tradeoffs})
    % \item Does the distributed futures interface actually provide both flexibility and high performance? (\cref{sec:eval:distributed-futures})
    % \item Are the components we require from the system both necessary and sufficient for \sys to achieve good performance? (\cref{sec:eval:mb})
    % \item \lsf{if time + able to fit into rest of eval: justify choice of decoupling architecture by showing it's easy to swap out storage backends and the application doesn't even have to be aware of the storage backend (S3 vs. disk); contrast w/ Magnet, which was specifically optimized for disk}
\end{itemize}

\begin{figure*}[t]
  \centering
  \begin{subfigure}[t]{0.4\textwidth}
      \centering
      \includegraphics[height=1.25in]{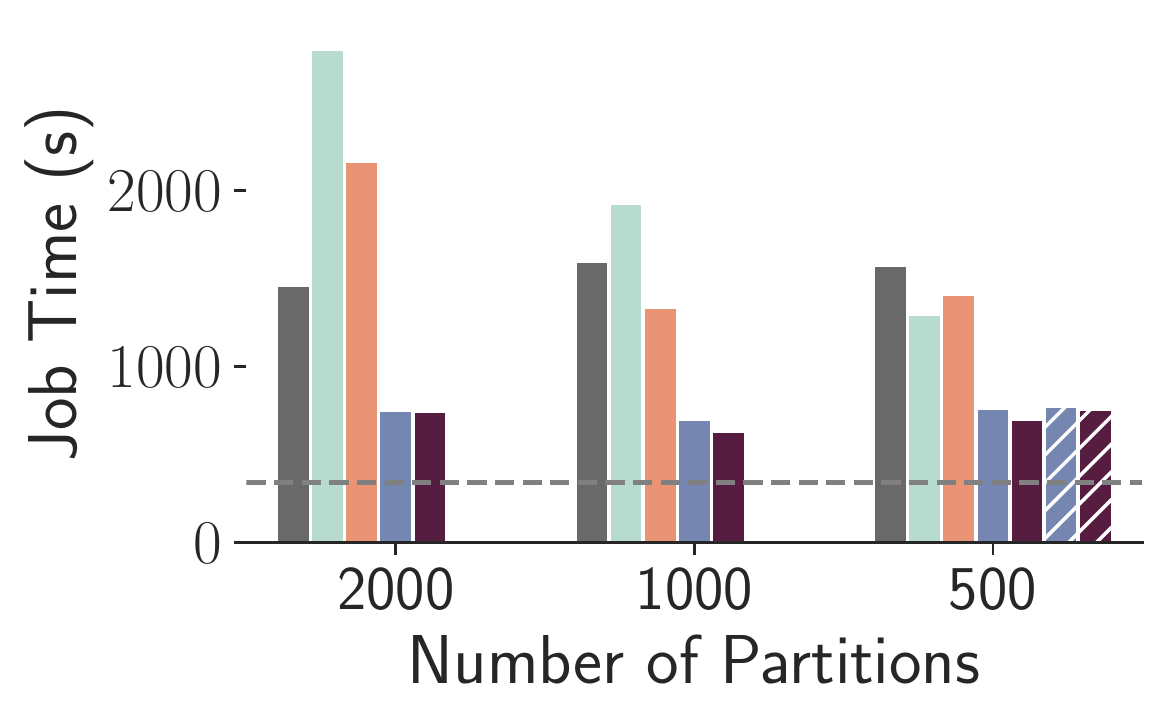}
      \caption{1\,TB sort on 10 HDD nodes.}
      \label{fig:shuffle-comparison-hdd}
  \end{subfigure}%
  \begin{subfigure}[t]{0.6\textwidth}
    \centering
    \includegraphics[height=1.25in]{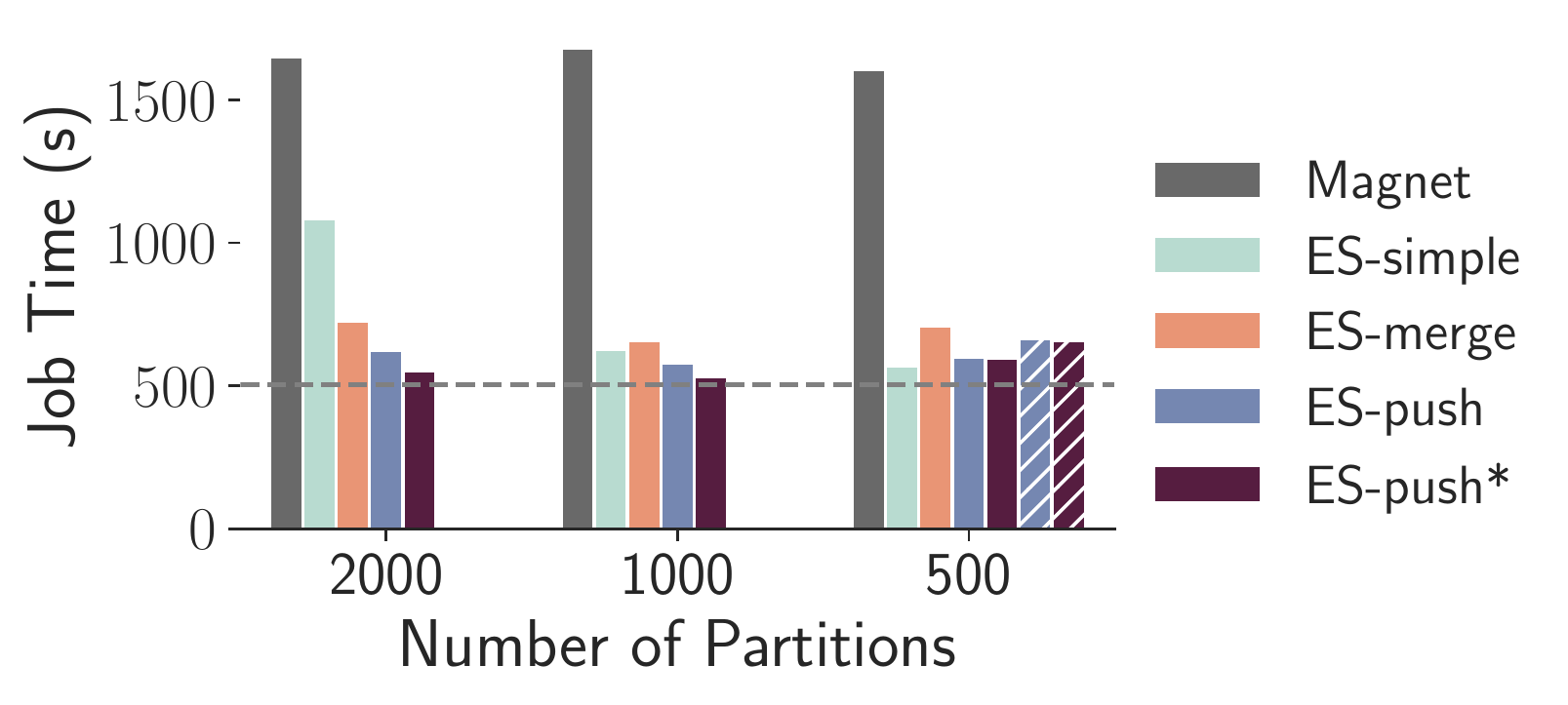}
    \caption{1\,TB sort on 10 SSD nodes. Semi-shaded bars are runs with failures (\cref{sec:eval:ft}).}
    \label{fig:shuffle-comparison-ssd}
  \end{subfigure} \\
  \begin{subfigure}[t]{0.47\textwidth}
      \centering
      \includegraphics[height=1.15in]{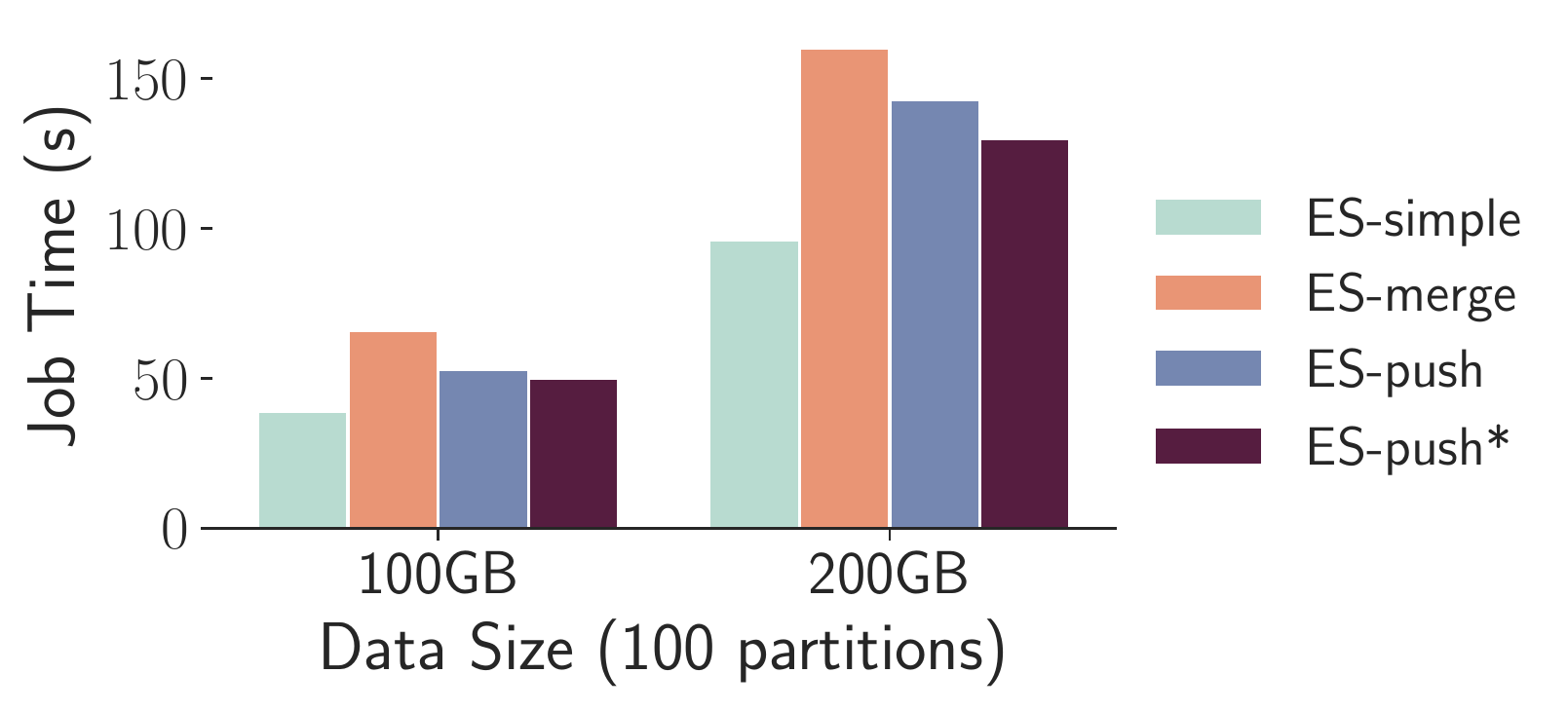}
      \caption{In-memory sort on 10 SSD nodes.}
      \label{fig:shuffle-comparison-small}
  \end{subfigure}%
  \begin{subfigure}[t]{0.53\textwidth}
      \centering
      \includegraphics[height=1in]{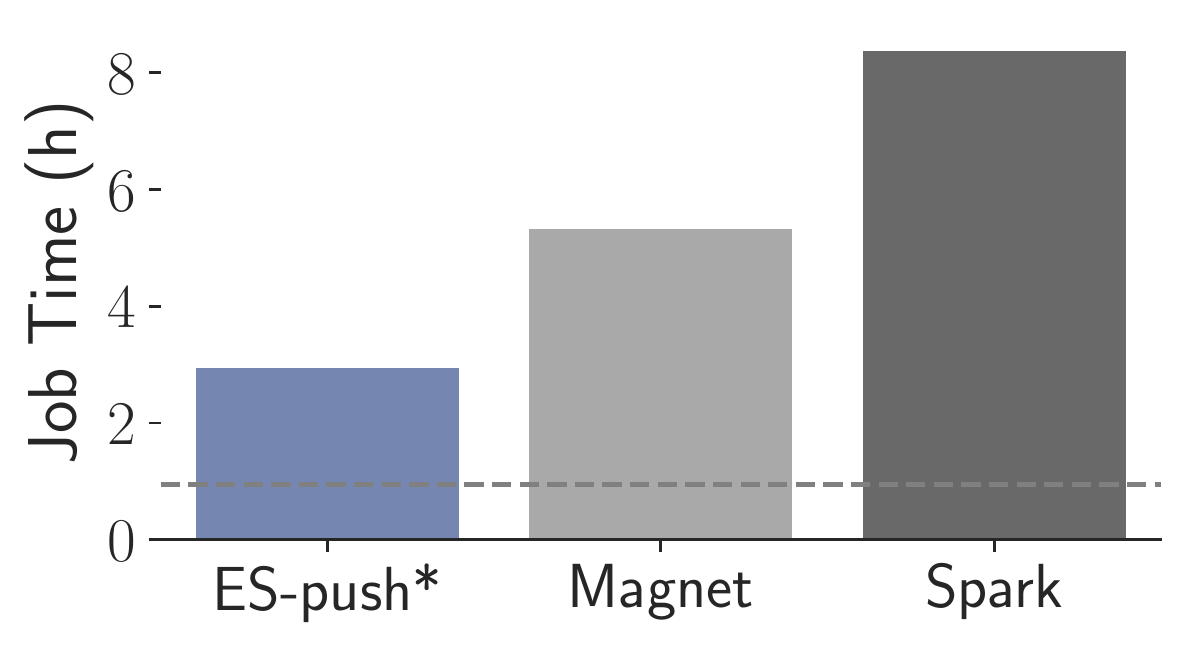}
      \caption{100\,TB on 100 HDD nodes.}
      \label{fig:shuffle-comparison-large}
  \end{subfigure}%
  \caption{Comparing job completion times on the Sort Benchmark. The dashed lines indicate the theoretical baseline (\cref{sec:eval:setup}). \textsf{\sys} is abbreviated as \textsf{\sysshort}.}
\end{figure*}

\subsection{Shuffle Performance}
\label{sec:eval:performance}

\subsubsection{Setup} \label{sec:eval:setup}
We create test environments on Amazon EC2 using VMs targeted at data warehouse use cases.
We test on a HDD cluster of \textsf{d3.2xlarge} instances (8\,CPU, 64\,GiB RAM, $6\times$ HDD, 1.1\,GB/s aggregate sequential throughput, 18K aggregate IOPS, 15\,Gbps network), and a SSD cluster of \textsf{i3.2xlarge} instances (8\,CPU, 61\,GiB RAM, NVMe SSD, 720\,MB/s throughput, 180K write IOPS, 10\,Gbps network).

\paragraph{Workload.}
We run the Sort Benchmark (a.k.a. TeraSort or CloudSort)~\cite{cloudsort}, as it is a common benchmark for testing raw shuffle system performance.
This benchmark requires sorting a synthetic dataset of configurable size, consisting of 100-byte records with 10-byte keys.

\paragraph{Baselines.}
We compare to the push-based shuffle service in Spark, a.k.a. \textsf{Magnet}, and a theoretical baseline.

\textsf{Magnet} is integrated into Spark in its \textsf{3.2.0} release as an external push-based shuffle service.
We run \textsf{Spark 3.2.0} on \textsf{Hadoop 3.3.1} with \textsf{Magnet} shuffle service enabled.
We disable compression of shuffle files according to the rules of TeraSort. This allows for a fair comparison in terms of total bytes of disk I/O.

For the theoretical baseline, we assume disk I/O is the bottleneck since empirically we find that disk I/O takes longer than networking and CPU processing in this benchmark.
The baseline is calculated by $T = 4D/B$, where $D$ is the total data size and $B$ is the aggregate disk bandwidth. $D$ is multiplied by 4 because each datum needs to be read twice and written twice, a theoretical minimum for external sort~\cite{themis}.

\paragraph{\sys variants.} 
We run \sys on \textsf{Ray 1.11.0}. We compare implementations of the following shuffle libraries:
\begin{itemize}
    \item \textsf{\sysshort-simple}, the simple shuffle variant (\cref{sec:shuffle:impl:naive}).
    \item \textsf{\sysshort-merge}, pull-based shuffle with pre-shuffle merge, similar to that in \textsf{Riffle} (\cref{sec:shuffle:impl:riffle}).
    \item \textsf{\sysshort-push}, push-based shuffle similar to \textsf{Magnet} (\cref{sec:shuffle:impl:magnet}).
    \item \textsf{\sysshort-push*}, push-based shuffle further optimized to reduce write amplification (\cref{sec:arch:example}).
\end{itemize}

\subsubsection{Performance Comparison of Shuffle Algorithms}
\label{sec:eval:perf-comparison}

\paragraph{Performance on HDD.}
\label{sec:eval:perf:hdd}
\Cref{fig:shuffle-comparison-hdd} shows the job completion times of \sys variants running 1\,TB sort on 10 HDD nodes.
\textsf{\sysshort-simple} shows the well-known scaling problem: performance degrades as the number of partitions increases, because the intermediate shuffle blocks become more in number and smaller in size both quadratically, quickly reaching disk IOPS limit.
The push-based shuffle variants (\textsf{\sysshort-push}, \textsf{-push*}) achieve better performance regardless of the number of partitions, thanks to the merging of shuffle blocks to increase disk I/O efficiency and the pipelining of disk and network I/O.
\textsf{\sysshort-merge} runs slower than \textsf{-simple} because merging the map output blocks incurs additional disk writes, which outweighs the I/O efficiency savings when the number of partitions is small, and only shows benefits when the number of partitions increases.
The \textsf{Magnet} baseline shows comparable performance.
In summary, \sys libraries demonstrate performance benefits that match the characteristics of their monolithic counterparts.

\paragraph{Performance on SSD.}
\label{sec:eval:perf:ssd}

\Cref{fig:shuffle-comparison-ssd} shows the same benchmark and variants running on the SSD cluster.
All variants of \sys outperform the \textsf{PBS} baseline, and display similar trends as on the HDD cluster.
The run times of the optimized versions of \sys are also close to the theoretical baseline.
Since the NVMe SSD supports much higher random IOPS, the I/O efficiency gains are less pronounced.

\paragraph{In-memory Performance.}
\label{sec:eval:perf:memory}

\Cref{fig:shuffle-comparison-small} shows that when data fits in memory, \textsf{\sysshort-simple} is actually the fastest algorithm compared to all other variants.
This is because the other algorithms create copies of data by merging them, triggering unnecessary disk spilling.
\textsf{Magnet} observes similar behavior for small datasets\footnote{\scriptsize The \textsf{Spark 3.3.1} documentation states: ``Currently [\textsf{Magnet}] is not well suited for jobs/queries which runs quickly dealing with lesser amount of shuffle data.''}.

\paragraph{Conclusion.}
\label{sec:eval:perf:conclusion}
These experiments show that the shuffle algorithms provided by \sys offer the same performance benefits as their monolithic counterparts.
Furthermore, the most performant shuffle algorithm depends on the data size and hardware configuration, and \sys offers the flexibility to choose the most suitable algorithm at the application level, without having to deploy multiple systems.

\subsubsection{Shuffle Scalability}
\label{sec:eval:large-scale}

To test performance at large scale, we run the Sort Benchmark on 100\,TB data with $50\,000 \times$ 2\,GB input partitions on a cluster of $100\times$ \textsf{d3.2xlarge} VMs.
For \sys, we run the \textsf{\sysshort-push*} variant since it is the most optimized for scale.
For baselines, we run both Spark's native shuffle (\textsf{Spark}) and its push-based shuffle service (\textsf{Magnet}).
We run both baselines with compression on because Spark without compression becomes unstable at this scale.

\Cref{fig:shuffle-comparison-large} shows the results.
\sys outperforms both native Spark shuffle and the push-based shuffle service \textsf{Magnet}, despite Spark's compression reducing total bytes spilled by 40\%.
\textsf{Magnet} improves shuffle performance by $1.6\times$ because it reduces random disk I/O.
\sys further improves performance over push-based Spark by $1.8\times$.
This difference comes from reduced write amplification in \textsf{\sysshort-push*}, which spills only the merged map outputs, while \textsf{Magnet} also spills the un-merged map outputs.
These additional writes provide faster failure recovery through improved durability, albeit at the cost of performance.
\sys allows the application to choose between these tradeoffs by using \textsf{\sysshort-push} vs. \textsf{\sysshort-push*}.

\begin{figure*}[t]
  \centering
  \begin{minipage}[t]{0.33\textwidth}
      \centering
      \includegraphics[height=1.1in]{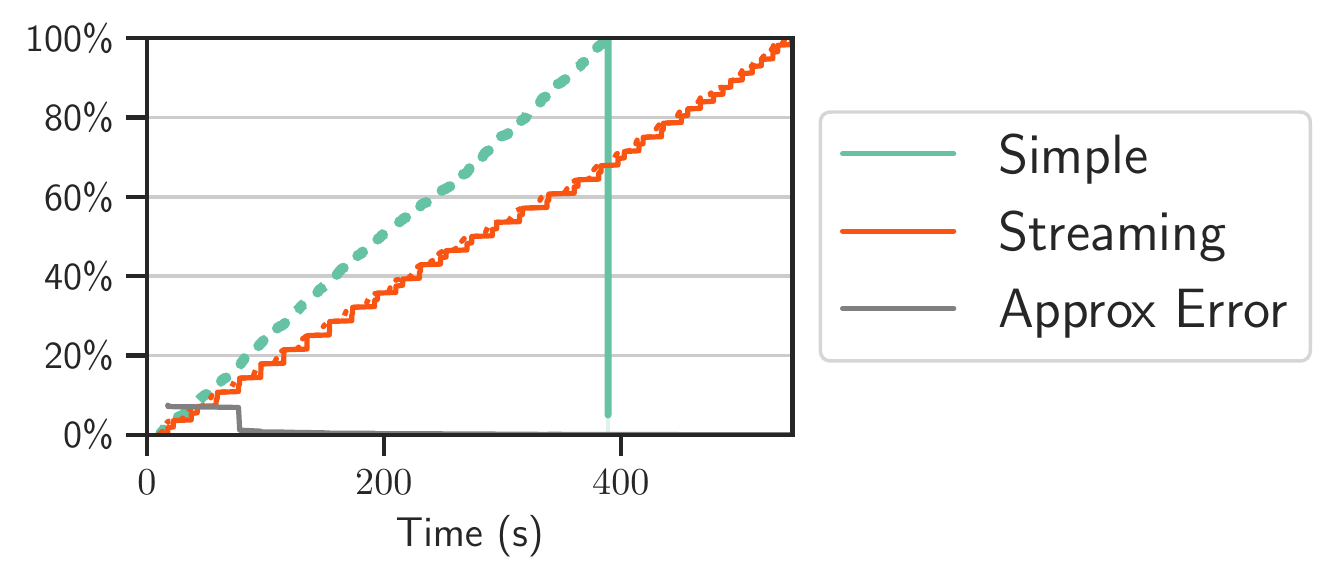}
      \caption{Online aggregation. Dotted lines show map progress; solid, reduce progress.}
      \label{fig:mpo}
  \end{minipage}%
  \hspace*{0.01\textwidth}
  \begin{minipage}[t]{0.32\textwidth}
      \centering
      \includegraphics[height=1.2in]{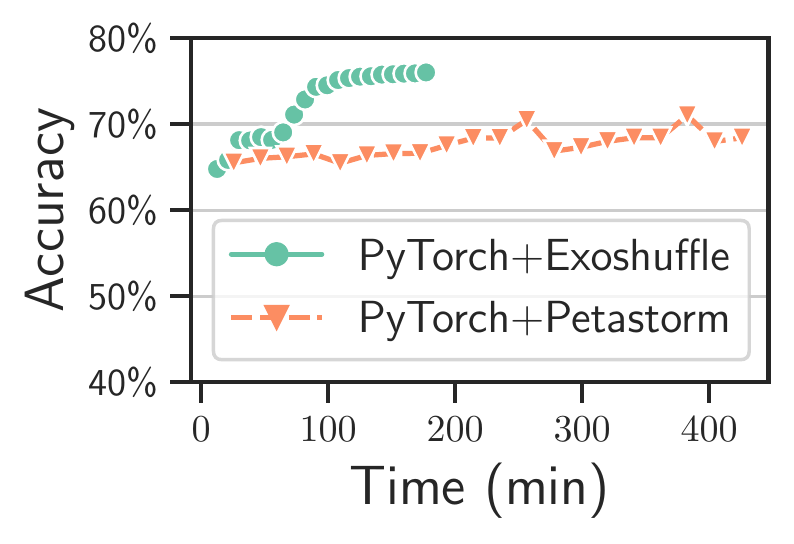}
      \caption{Single-node ML training for 20 epochs.}
      \label{fig:ludwig-single}
  \end{minipage}%
  \hspace*{0.01\textwidth}
  \begin{minipage}[t]{0.33\textwidth}
    \centering
    \includegraphics[height=1.2in]{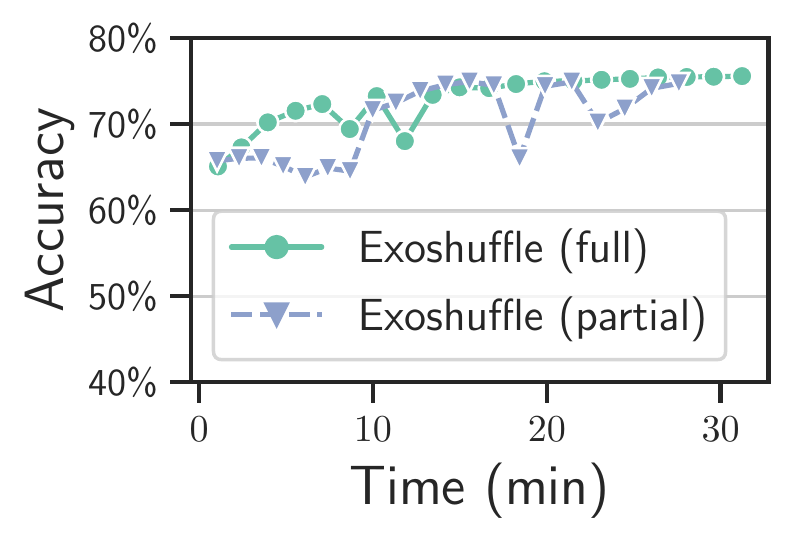}
    \caption{4-node, distributed ML training for 20 epochs.}
    \label{fig:ludwig-distributed}
  \end{minipage}%
\end{figure*}
\begin{figure*}[t]
  \centering
  \begin{minipage}[t]{0.48\textwidth}
      \centering
      \includegraphics[height=1.25in]{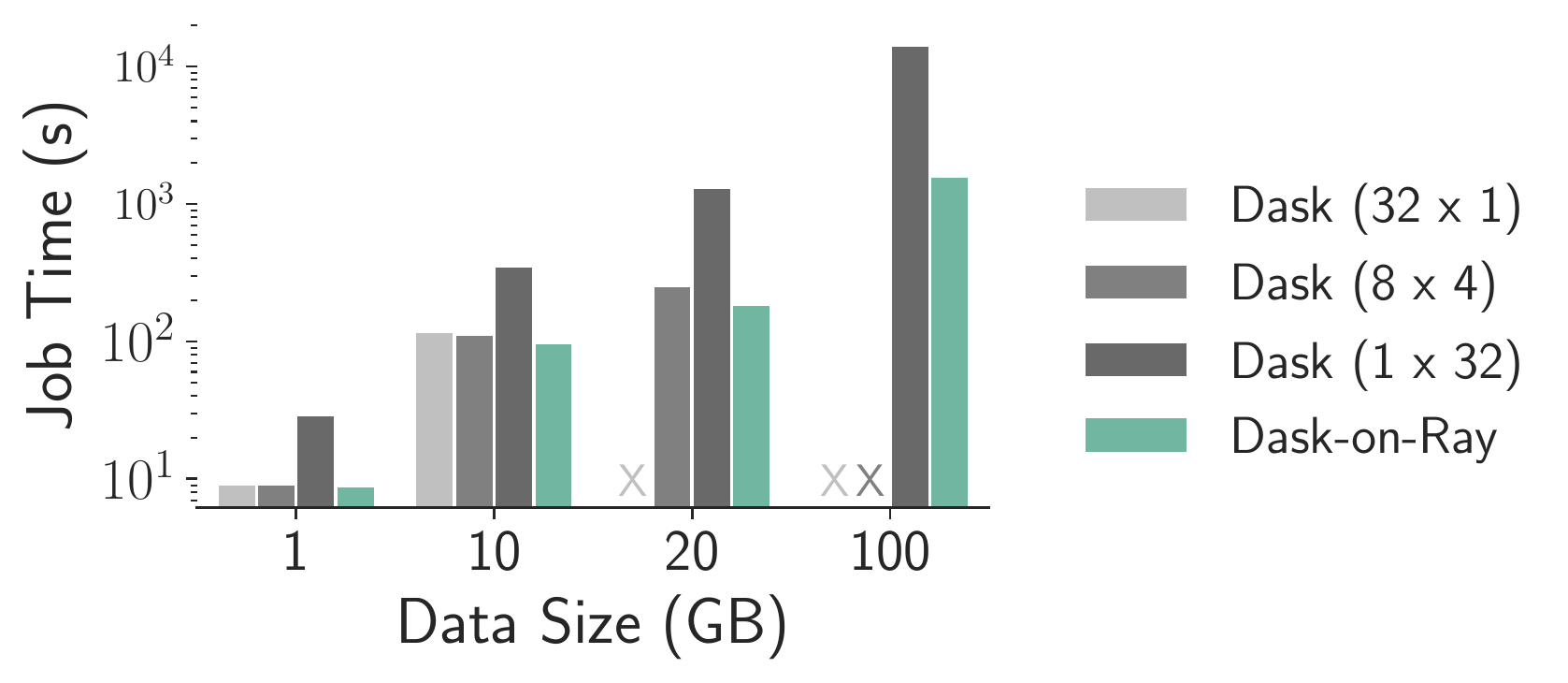}
      \caption{Comparing shuffle time in Dask and Ray. Legends show number of processes $\times$ threads.}
      \label{fig:dask_vs_ray}
  \end{minipage}%
  \hspace*{0.01\textwidth}
  \begin{minipage}[t]{0.48\textwidth}
      \centering
      \includegraphics[height=1.25in]{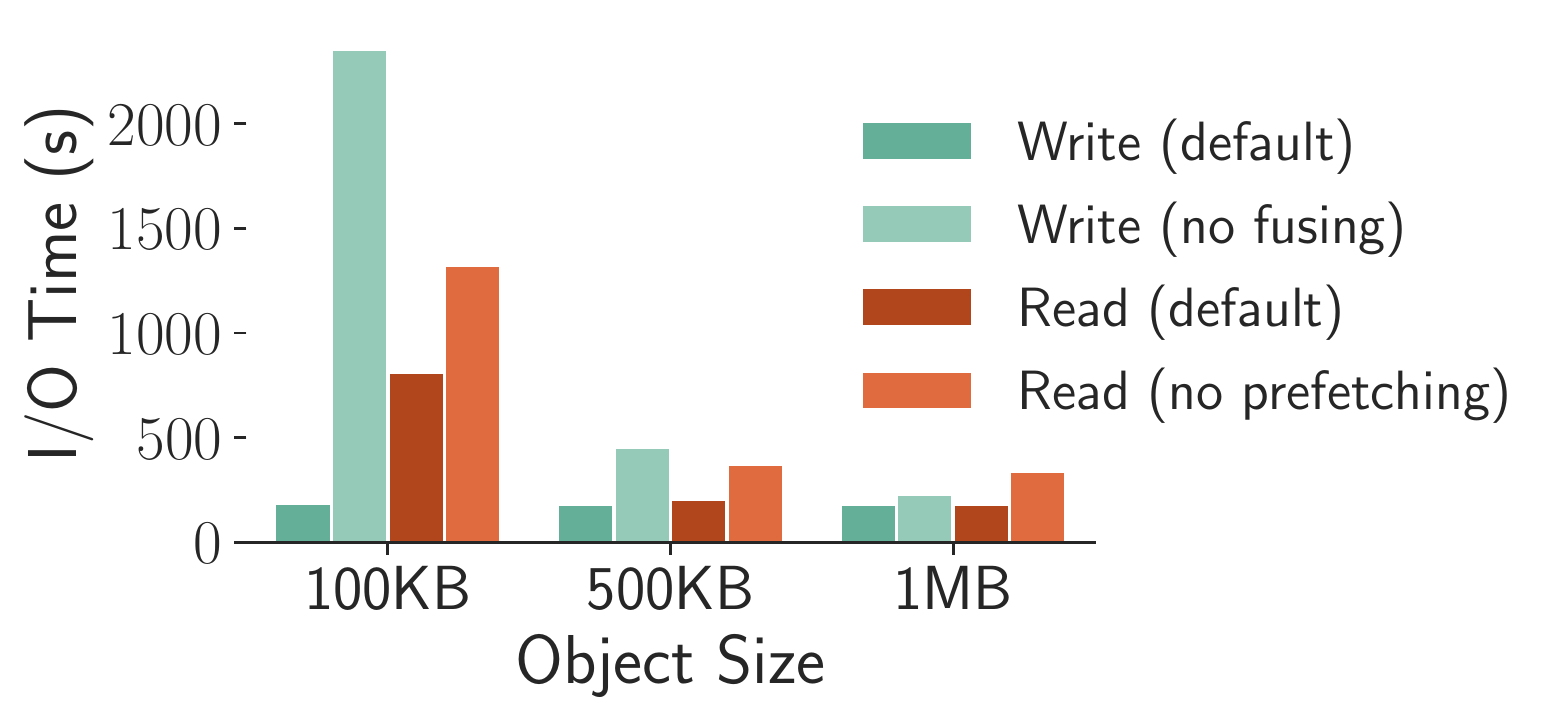}
      \caption{Effect of I/O optimizations in Ray.}
      \label{fig:mb-all}
  \end{minipage}%
\end{figure*}

\begin{table}[t]
\small
  \begin{tabular}{lll}
    \toprule
    Shuffle Algorithm
    &
    System LoC
    &
    \sys LoC
    \\
    \midrule
    Simple (\cref{sec:shuffle:impl:naive})
    &
    2600 (\textsf{Spark}\tablefootnote{\scriptsize Total lines of code in \texttt{\scriptsize org.apache.spark.shuffle}.})
    &
    \textbf{215}
    \\
    Pre-shuffle merge (\cref{sec:shuffle:impl:riffle})
    &
    4000 (\textsf{Riffle}\tablefootnote{\scriptsize As reported by Zhang et al.~\cite{riffle}})
    &
    \textbf{265}
    \\
    Push-based shuffle (\cref{sec:shuffle:impl:magnet})
    &
    6700 (\textsf{Magnet}\tablefootnote{\scriptsize Total added lines in \url{https://github.com/apache/spark/pull/29808/files}.})
    &
    \textbf{256} \\
    \hspace{1em} with pipelining (\cref{sec:arch:example})
    & --
    &
    \textbf{256} \\
  \bottomrule
\end{tabular}
  \caption{Approximate lines of code for implementing shuffle algorithms in \sys versus in specialized shuffle systems.
}
  \label{tab:loc}
\end{table}

\subsection{Implementation Complexity}
\label{sec:eval:complexity}

In \sys, shuffle is expressed as application-level programs.
\Cref{tab:loc} compares the amount of code of several monolithic shuffle systems with the lines of code needed to implement the corresponding shuffle algorithms in \sys.
\sys libraries may not provide all the production features of the monolithic counterparts, but many shuffle optimizations can be implemented in \sys with an order of magnitude less code, while keeping the same performance benefits.
By offering shuffle as a library, \sys also allows applications to choose the best shuffle implementation at run time without deploying multiple systems.

\subsection{Shuffle Applications}
\label{sec:eval:app}

Next, we show how \sys can extend distributed shuffle support for a broader set of applications.

\subsubsection{CloudSort}
\label{sec:eval:cloudsort}

The CloudSort competition~\cite{cloudsort} calls for the most cost-efficient way to sort 100\,TB of data on the public cloud.
We ran \sys-CloudSort on a cluster of $40\times$ \textsf{i4i.4xlarge} nodes with input and output data stored on Amazon S3, and set a new world record of \$0.97/TB~\cite{exoshuffle-cloudsort}. This is 33\% more cost-efficient than the previous world record set in 2016.
The previous entry used a heavily modified version of Spark for the CloudSort workload~\cite{NADSort}.
In contrast, \sys-CloudSort is only hundreds of lines of application code running on a release version of Ray.

To account for the fact that the cloud hardware costs have lowered since 2016, we take the setup from the previous record-winning entry and look up its cost on today's Alibaba Cloud.
\Cref{tab:cloudsort} shows that the same amount of cloud resources would cost \$115 today.
Still, \sys-CloudSort achieves another 15\% cost reduction beyond this result.
We calculate another theoretical baseline of simply shuffling 100\,TB data through the AWS network (without sorting), which would cost \$74.
This puts our record within 31\% of the theoretical limit.
This result demonstrates that \sys can achieve state-of-the-art performance and cost-efficiency for large-scale shuffle.

\begin{table}[ht]
\small
\begin{tabular}{ll}
    \toprule
    System
    &
    Cost
    \\
    \midrule
    NADSort (2016)
    &
    \$1.44/TB
    \\
    NADSort (2022, extrapolated)
    &
    \$1.15/TB
    \\
    \sys-CloudSort (2022)
    &
    \textbf{\$0.97/TB} \\
  \bottomrule
\end{tabular}
\caption{CloudSort costs over years.}
\label{tab:cloudsort}
\end{table}

\subsubsection{Online Aggregation with Streaming Shuffle}
\label{sec:eval:mpo}

We use a 1\,TB dataset containing 6 months of hourly page view statistics on Wikipedia.
We run an aggregation to get the ranking of the top pages by language on $10\times$ \textsf{r6i.2xlarge} nodes with data loaded from S3.
\Cref{fig:mpo} shows the difference between regular and streaming shuffle.
The streaming shuffle takes $1.4\times$ longer to run in total due to the extra computation needed to produce partial results.
However, with streaming shuffle, the user can get partial aggregation results within 8\% error\footnote{\scriptsize Error is computed using the KL-divergence $D_{KL} = \sum{p \log (p/\hat{p})}$ where $p$ is the true statistic and $\hat{p}$ is the sample statistic.} of the final result in 18 seconds, $22\times$ faster than regular shuffle.
\sys makes it easy to switch between \texttt{simple\_shuffle} and \texttt{streaming\_shuffle} to choose between partial result latency and total query run time.

\subsubsection{Distributed ML Training}
\label{sec:eval:ml-training}

Many distributed training frameworks already run on top of Ray.
By offering \sys as a library, we enable these workloads to leverage scalable shuffle.
We demonstrate \sys's ability to support fine-grained pipelining for ML training using the Ludwig framework~\cite{ludwig} to train a deep classification model TabNet on the HIGGS dataset (7.5\,GB).
Ludwig integrates ML data loaders with the PyTorch training framework~\cite{pytorch}.
Efficient training requires randomly shuffling the data per epoch before sending it into the GPU for training.

We first run the ML training on a single \textsf{g4dn.4xlarge} instance. We compare two versions of Ludwig:
Ludwig~0.4.0 uses Petastorm~\cite{petastorm}, which prefetches data in batches into a per-process memory buffer and performs random shuffle in the buffer.
\revision{
This approach makes the shuffle window size limited by the memory buffer size (\cref{sec:background:ml}).
}
In this experiment, we set the shuffle window size to 9\% of the total data size to avoid OOM errors.
In comparison, Ludwig~0.4.1 uses \sys offered through Ray Data~\cite{ray.data}.
It pipelines data loading and shuffling with GPU training (\cref{fig:shuffle-dags:ml}), and supports full shuffle across loading batches by storing data in the shared-memory object store.
\Cref{fig:ludwig-single} shows that model training with \sys is $2.4\times$ faster end-to-end thanks to the fine-grained pipelining.
The model also converges faster per-epoch and to a higher accuracy, because \sys performs complete random shuffling between epochs, whereas Petastorm's random shuffle is limited to subsets of the data.

Next, we run the training on 4 \textsf{g4dn.xlarge} nodes to show the distributed shuffle performance.
Ludwig 0.4.$x$ has known bugs with distributed training, so we could not compare Petastorm with \sys.
Instead, we use the latest Ludwig 0.6.0 and compares two shuffle strategies with the \sys-based data loader: full shuffle (the default) and partial shuffle.
For partial shuffle, we emulate the Petastorm behavior and perform random shuffling only in each in-memory batch.
\Cref{fig:ludwig-distributed} shows that per-epoch time is slightly faster with partial shuffle, since it is fully local, but the convergence accuracy is slightly lower because of the less random shuffling of training data.
This example demonstrates that \sys gives the developer the flexibility to choose the best shuffle strategy based on their training needs, while providing high-throughput data loading and shuffling.

\subsection{System Microbenchmarks} \label{sec:eval:mb}

The \sys architecture requires high-performance components from the distributed futures system to deliver good performance.
In this section, we study the the impact of these system components on shuffle performance.

\subsubsection{Shared-Memory Object Store}
\label{sec:eval:dask}

We study the effect of a shared-memory object store that decouples objects from executors by comparing Dask and Ray.
Dask and Ray are both distributed futures systems, but they differ in architecture.
Ray uses a shared-memory object store that is shared by multiple executor processes on the same node~(\cref{sec:arch:system:obj-store}).
Dask stores objects in executor memory and requires the user to choose between multiprocessing and multithreading.
With multithreading, multiple Dask executor threads share data in a heap-memory object store, but the Python Global Interpreter Lock can severely limit parallelism.
Dask in multiprocessing mode avoids this issue but uses one object store per worker process, so objects must be copied between workers on the same node.
Thus, the lack of a shared-memory object store results in either reduced parallelism (multithreading) or high overhead for sharing objects (multiprocessing). It is also less robust as objects are vulnerable to executor failures.

We study these differences by running the same Dask task graph on Dask and Ray backends~\cite{dask-on-ray}.
\Cref{fig:dask_vs_ray} shows dataframe sorting performance on a single node (32 CPU, 244\,GB RAM, 100 partitions).
For Dask, we vary the number of executor processes and threads to show the tradeoff between memory usage and parallelism.
Ray requires no configuration and uses 32 executor processes, 1 per CPU.

On small data sizes, Dask with multiprocessing achieves about the same performance as Ray, but it is 3$\times$ slower with multithreading due to reduced parallelism.
On larger data sizes, Dask with multiprocessing fails due to high memory pressure from extra object copies.
Meanwhile, Ray's shared-memory object store enables better stability and lower run time on all data sizes.

\subsubsection{Small I/O Mitigations}

Ray implements two system-level optimizations for mitigating the small I/O problem: fusing writes of spilled objects to avoid small disk I/O, and prefetching task arguments to hide network and disk latency (\cref{sec:arch:memory:spill}).
To show the impact of these optimizations, we run a single-node microbenchmark that creates 16\,GB total objects in a 1\,GB object store, forces them to spill to disk, then restores the objects from disk.
We use object sizes ranging from 100\,KB to 1\,MB, as these are comparable to the shuffle block sizes. We use a \textsf{sc1} HDD disk since the disk I/O bottleneck is more pronounced on slower storage.

\paragraph{Fusing Writes.}
Ray fuses objects into at least 100\,MB files then writes them to disk.
\Cref{fig:mb-all} shows the total run time stays constant across object sizes with default fusing.
When fusing is off, the run time is 25\% slower for 1\,MB objects, and up to $12\times$ slower when spilling 100\,KB objects.

\paragraph{Prefetching Task Arguments.}
Ray prefetches task arguments in a pipelined manner so that arguments are ready on a worker by the time the task is scheduled.
\Cref{fig:mb-all} shows that pipelined fetching of task arguments reduces the run time by 60--80\%, comparing with a baseline implementation that only starts fetching objects after the task is scheduled.

\subsubsection{Fault Tolerance} \label{sec:eval:ft}
To test fault recovery, we fail and restart a random worker node 30 seconds after the start of the run.
This results in both executor failure and data loss, as the worker's local object store is also lost.
In all cases, we rely on the distributed futures system to re-execute any lost tasks and to reconstruct any lost objects.
Lineage reconstruction (\cref{sec:arch:ft}) minimizes interruption time during worker failures.
\Cref{fig:shuffle-comparison-hdd,fig:shuffle-comparison-ssd} show run times with failures indicated with semi-shaded bars.
For \textsf{\sysshort-simple} and \textsf{-merge}, a known bug in Ray currently prevents fault recovery from completing.
For \textsf{\sysshort-push} and \textsf{-push*}, recovering from a worker failure adds 20--50 seconds to the job completion time.
The system uses this time to detect node failures and re-execute tasks to reconstruct lost objects.

\section{Related Work} \label{sec:related}

\paragraph{Shuffle in Data Processing Systems.}

Many solutions to shuffle have been proposed~\cite{sailfish,ishuffle,jvm-shuffle,improving-shuffle-hadoop,tritonsort,themis} since MapReduce~\cite{mapreduce} and Hadoop~\cite{hadoop}, with a focus on optimizing disk I/O and pipelining.
Sailfish~\cite{sailfish} is a notable example deployed at Yahoo which depends on a modified filesystem to batch disk I/O.
Many recent shuffle systems have been built in industry for large-scale use cases~\cite{riffle,cosco,magnet,zeus}, but few have been open-sourced.
Today's cloud providers often offer managed shuffle services~\cite{alibaba-emr,google-dataflow-shuffle,awsglueshuffle}.
However these are tightly integrated with proprietary cloud data services and are not accessible by other shuffle applications.

\paragraph{Hardware Environments.}
Hardware typically poses a range of constraints on shuffle design. For example, compute and memory may be either disaggregated~\cite{riffle,cosco,locus} or colocated \cite{sailfish,sparkpetabyteblog,magnet}.
Disk constraints also affect system design, e.g., SSDs provide better random IOPS than HDDs but wear out more quickly.
Many existing shuffle systems have been motivated by such hardware differences.
In \sys, because the distributed futures API abstracts block management, a shuffle developer can plug in different storage backends and optimize shuffle at the application level.

\paragraph{Other Shuffle Applications.}

Machine learning research~\cite{randomreshuffling,sgd-shuffle} shows that SGD-based model training benefits from random shuffling of the training dataset.
Both TensorFlow~\cite{tfdata} and PyTorch~\cite{torchdata} have built specialized systems designed specifically to pipeline data loading with ML training.
These data loaders, in addition to Petastorm~\cite{petastorm}, support distributed data loading and random shuffling but shuffling is limited to a local buffer capped by worker memory~(\cref{sec:eval:ml-training}).

Dataframes~\cite{pandas,dask,modin} are another class of applications in data science that depend on shuffle for operations such as group-by.
While systems like Dask~\cite{dask} and Spark~\cite{spark} provide distributed dataframes, developers continue to build new engines that optimize for specific application scenarios, such as multi-core~\cite{polars}, out-of-core performance~\cite{vaex}, or supporting SQL~\cite{datafusion}.
These new dataframe libraries, along with new embedded query engines such as DuckDB~\cite{duckdb} and Velox~\cite{velox}, can directly use \sys to support distributed query processing.

\paragraph{Distributed Programming Abstractions for Shuffle.}

CIEL~\cite{ciel} is the first to propose using distributed futures to express iterative distributed dataflow programs, including MapReduce.
Its implementation does not include features critical to large-scale shuffle performance, including intra-node parallelism, in-memory object storage, and automatic garbage collection~\cite{murray2012distributed}.
Dask~\cite{dask} is another distributed futures-based system that has trouble scaling shuffle due to the lack of shared-memory objects (\cref{sec:eval:dask}).
While we build on Ray's design, such as a shared-memory object store~\cite{ray} and lineage reconstruction~\cite{ownership}, previous versions are not sufficient to support large-scale shuffle as they do not include spilling to disk or pipelining between execution and I/O.
Thus, while others have implemented shuffle on distributed futures before, ours is the first that we know of to reach the scale, performance, and reliability of monolithic shuffle systems.

Serverless functions, as used in Locus~\cite{locus}, are one alternative to distributed futures.
While Locus leverages an existing serverless cache and persistent storage, it still must manage block movement manually.
In contrast, distributed futures abstract block management in full and manage execution, memory, and disk collectively on each node.

\revision{
Hoplite~\cite{hoplite} shows that it is possible to provide a high-performance and fault-tolerant collective communication layer on top of distributed futures, supporting operations such as scatter, gather, and reduce.
Shuffle in MapReduce-like systems is a more challenging problem because it involves scheduling arbitrary compute tasks along with all-to-all communication.
In this work we show that a distributed futures system can support shuffle at TB+ scale and provide competitive performance and reliability.
}

\section{Discussion}
\label{sec:discussion}

\paragraph{Extensible Architectures.}

The decoupling of control and data planes in software-defined networking~\cite{openflow} has led to great innovations in the past two decades~\cite{roadtosdn}.
Operating systems research also advocates for extensible architectures to build OS kernels, such as microkernels~\cite{mach} and exokernels~\cite{exokernel}.
We hope our work can drive more innovations in shuffle designs and applications through an extensible architecture for distributed shuffle.

\paragraph{Distributed Futures.}
Distributed futures are rising in popularity due to their ease of use and flexibility~\cite{ciel,ray,hotos-rpc}.
However, the question of flexibility versus performance remains.
Large-scale shuffle is one of the most challenging problems in big data processing, inspiring years of work.
By showing that large-scale shuffle is possible on a generic and flexible distributed futures system, we hope to show that other complex applications can be built on this framework, too.

\paragraph{Limitations.}
The ability to specify arbitrary tasks and objects with distributed futures is the key to its flexibility, but it is also the primary obstacle to performance.
The system assumes that each task is independent for generality and stores metadata separately for each task and object.
In contrast, monolithic shuffle systems have semantic information and can share metadata for tasks and objects in the same stage.
Currently metadata overhead is the main limitation to executing \sys at larger scales. We plan to address this in the future by ``collapsing'' shared metadata, i.e., keeping one metadata entry for multiple outputs of a task.

Architecturally, the primary limitation in \sys is the fact that an object must be loaded in its entirety into the local object store before it can be read~(\cref{sec:arch:system:obj-store}).
Generators allow tasks to ``stream'' large outputs by breaking them into many smaller physical objects; future improvements include the described metadata optimizations and/or introducing APIs to stream objects larger than the object store, similar to Ciel~\cite{ciel}.
Another limitation is in scheduling.
Currently the distributed futures system may require hints from the shuffle library to determine which tasks should be executed concurrently and where to place tasks~(\cref{sec:arch:scheduling}).
A more sophisticated scheduler may be able to determine these automatically.

Finally, \sys does not yet address the problem of providing a single shuffle solution that can meet the requirements of all applications.
Doing so would require automatically picking the best shuffle algorithm and parameters based on application, environment, and run-time information.
Instead, we focus on the problem of shuffle \emph{evolvability}, a necessary step towards this overarching goal.

\section{Conclusion}
\label{sec:conclusion}

There is a longstanding tension between performance and flexibility in designing abstractions for distributed computing.
Monolithic shuffle systems sacrifice flexibility in the name of performance: they must essentially rebuild shuffle from scratch to handle varying application scenarios. %: coordination, block management, and fault tolerance.
In this work, we show that this need not be the case, by demonstrating an extensible architecture with a distributed futures system that makes it possible build efficient, flexible, and portable shuffle.

This paper does not raise ethical issues.
\section*{Acknowledgement}

We thank the SIGCOMM reviewers and our shepherd Yiting Xia for their valuable feedback.
We also thank members of the Sky Computing Lab at UC Berkeley for all the insightful discussion around this work.
We thank Anyscale for providing the cloud resources necessary for completing the experiments in this work.
This work is in part supported by NSF CISE Expeditions Award CCF1730628, and gifts from Astronomer, Google, IBM, Intel, Lacework, Microsoft, Mohamed Bin Zayed University of Artificial Intelligence, Nexla, Samsung SDS, Uber, and VMware.

\clearpage

\bibliographystyle{ACM-Reference-Format}
\bibliography{_main}

\end{document}